\documentclass[12pt,english]{article}
\pdfoutput=1
\usepackage[T1]{fontenc}
\usepackage{amssymb}
\usepackage{amsmath}
\usepackage{cite}
\usepackage{epsfig}
\usepackage[unicode=true,
 bookmarks=true,bookmarksnumbered=false,bookmarksopen=false,
 breaklinks=false,pdfborder={0 0 1},backref=false,colorlinks=true]
 {hyperref}
\usepackage[hang,flushmargin]{footmisc}
\usepackage{relsize}
\usepackage{color}

\usepackage[latin1]{inputenc}
\usepackage{amsmath}
\usepackage{amsfonts}
\usepackage{amssymb}
\usepackage{graphicx}

\usepackage{tikz,environ}
\usetikzlibrary{decorations}
\usetikzlibrary{arrows}
\usetikzlibrary{decorations.markings}
\tikzset{
	hard/.style={postaction={decorate},
		line width=0.5mm
	},
		soft/.style={postaction={decorate},
		line width=0.5mm, dashed
	},
	momentum/.style={postaction={decorate},
	line width=0.5mm,
	color=gray,
	decoration={
    markings,
    mark=at position 0.8 with {\arrow{stealth}}}
    },
	hardarrow/.style={postaction={decorate},
	line width=0.5mm,
	decoration={
    markings,
    mark=at position 0.6 with {\arrow{stealth}}}},
}

\usepackage{pifont}% http://ctan.org/pkg/pifont

\makeatletter
\def\simgt{\mathrel{\lower2.5pt\vbox{\lineskip=0pt\baselineskip=0pt
           \hbox{$>$}\hbox{$\sim$}}}}
\def\simlt{\mathrel{\lower2.5pt\vbox{\lineskip=0pt\baselineskip=0pt
           \hbox{$<$}\hbox{$\sim$}}}}
\makeatother

\newcommand{\Lagr}{L}

\newcommand{\Amp}[1]{A_{#1}}
\newcommand{\Ampu}[1]{A^{#1}}

\newcommand{\J}{\mathcal{J}}

\newcommand{\be}{\begin{equation}}
\newcommand{\ee}{\end{equation}}
\newcommand{\eq}[2]{\be\begin{aligned}#1 \label{#2}\end{aligned}\ee}

\newcommand{\Sec}[1]{Sec.~\ref{#1}}

\newcommand{\Eq}[1]{Eq.~\eqref{#1}}

\newcommand{\cI}{I}
\newcommand{\cJ}{J}
\newcommand{\cK}{K}
\newcommand{\cL}{L}
\newcommand{\cM}{M}
\newcommand{\cP}{P}
\newcommand{\cQ}{Q}
\newcommand{\cR}{R}
\newcommand{\cS}{S}

\newcommand{\Kill}{{\cal K}}
\newcommand{\XKill}{{\cal X}}
\newcommand{\TKill}{{\cal T}}

\newcommand{\dil}{\xi}

\newcommand{\fI}{i}
\newcommand{\fJ}{j}
\newcommand{\fK}{k}
\newcommand{\fL}{l}

\begin{document}

\hypersetup{citecolor=blue,linkcolor=blue,urlcolor=blue}

\interfootnotelinepenalty=10000
\baselineskip=18pt

\hfill CALT-TH-2021-038

\vspace{2cm}
\thispagestyle{empty}
\begin{center}
{\LARGE \bf
Geometric Soft Theorems \\
}
\bigskip\vspace{1cm}{
{\large Clifford Cheung, Andreas Helset, Julio Parra-Martinez}
} \\[7mm]
 {\it Walter Burke Institute for Theoretical Physics,\\[-1mm]
    California Institute of Technology, Pasadena, CA 91125 } 
 \end{center}

\bigskip
\centerline{\large\bf Abstract}
\begin{quote} \small

We derive a universal soft theorem for every scattering amplitude with at least one massless particle in an arbitrary theory of scalars.  Our results follow from the geometry of field space and are valid for any choice of mass spectrum, potential terms, and higher-derivative interactions.  For a vanishing potential, the soft limit of every amplitude is equal to the field-space covariant derivative of an amplitude with one fewer particle.   Furthermore, the Adler zero and the dilaton soft theorem are special cases of our results.  We also discuss more exotic scenarios in which the soft limit is non-trivial but still universal.   Last but not least, we derive new theorems for multiple-soft limits which directly probe the field-space curvature, as well as on-shell recursion relations applicable to two-derivative scalar field theories exhibiting no symmetries whatsoever.

\end{quote}

\setcounter{footnote}{0}

\newpage

\tableofcontents
\newpage

\section{Introduction}

The long distance dynamics of certain physical systems exhibit universal behaviors.  Typically, these infrared structures are dictated by the underlying symmetries of the theory.  For example, scattering amplitudes in gauge theory and gravity conform to soft theorems  which are direct consequences of charge and energy-momentum conservation, respectively \cite{Low:1958sn,Weinberg:1965nx, Burnett:1967km, Bern:2014vva}.  Another notable instance is the Adler zero \cite{Adler:1964um}, which mandates the vanishing of soft pion amplitudes due to non-linearly realized, spontaneously broken chiral symmetries.

These known examples suggest that universal soft behavior should only be expected of theories exhibiting some incarnation of enhanced symmetry.  In this paper we show that this intuition is incorrect.  In particular, we derive a soft theorem that applies to a {\it completely general theory} of interacting scalar fields.  Our results are applicable in the presence of masses, potential interactions, and higher-derivative couplings, sans any underlying linear or non-linear symmetries. 

The key strategy in our analysis is to interpret scattering amplitudes not merely as functions of external kinematics, but also as functions of the vacuum expectation values (VEVs) of scalar fields.   As is well-known, every scalar field theory is endowed with a notion of geometry in which the scalars parameterize coordinates of an underlying field-space manifold \cite{Meetz:1969as,Honerkamp:1971sh,Honerkamp:1971xtx, Ecker:1971xko,Alvarez-Gaume:1981exa,Alvarez-Gaume:1981exv,Boulware:1981ns,Howe:1986vm}.   Furthermore, since on-shell scattering amplitudes are physical quantities, they are necessarily functions of the corresponding geometric invariants \cite{Dixon:1989fj,Alonso:2015fsp,Alonso:2016oah,Nagai:2019tgi,Cohen:2021ucp}.

As we will prove, the soft limit $q\rightarrow 0$ for a massless scalar of a tree-level $(n+1)$-particle amplitude is related to the tree-level $n$-particle amplitude  by a {\it geometric soft theorem} of the schematic form,
\eq{
	\lim_{q\rightarrow 0} \Amp{n+1}  \sim \left(\nabla +  \frac{\nabla m^2}{p^2-m^2} \right) \Amp{n} \, ,
}{eq:soft_thm_schematic}
where $\nabla$ is a covariant derivative in {\it field space} and $p$ and $m$ are the momentum and mass of an exchanged state. The geometric soft theorem has an elegant and intuitive physical interpretation.  The soft limit of the massless particle is equivalent to an infinitesimal shift of the VEV of the corresponding scalar.   From this perspective, the first term in \Eq{eq:soft_thm_schematic}  corresponds to the variation of interaction vertices and internal propagators with respect to the VEV.   The second term arises from the variation of amputated external propagators---in particular the masses appearing therein---with respect to the VEV.

Even though \Eq{eq:soft_thm_schematic} holds in general, it is illuminating to consider its application to systems of particular interest.  For instance, we study examples with an arbitrary potential, as well as two- or higher-derivative interactions.  Our results also hold for any theory with linear or non-linear symmetries.  The latter case describes the dynamics of spontaneous symmetry breaking, where the potential is absent and the field-space manifold is a coset.  When the coset manifold exhibits isometries, the right-hand side of the soft theorem is in general non-zero but is directly related to the associated Killing vectors.  If the coset is also symmetric, then the right-hand side is zero, thus proving the Adler zero condition.  We also show how the dilaton soft theorem is a special case of \Eq{eq:soft_thm_schematic}.

As another application, we derive new theorems describing amplitudes when multiple particles are taken soft, either consecutively or simultaneously, in a general theory with vanishing potential.   We describe how these multiple-soft theorems characterize the local curvature of field space.   In addition, we derive on-shell recursion relations that crucially leverage the geometric soft theorem.  Remarkably, many theories that are lacking in any enhanced symmetry properties are still on-shell constructible via these methods.

The remainder of this paper is structured as follows. In \Sec{sec:geometry}, we briefly describe the geometry of scalar field space.  The bulk of this discussion is review, but we do present some new results, {\it e.g.}~how the wavefunction normalization factors in the Lehmann, Symanzik and Zimmermann (LSZ) reduction formula \cite{Lehmann:1954rq} have the geometric interpretation as tetrads that toggle between fields and physical scattering states.
Afterwards, in \Sec{sec:theorem} we present an explicit definition of the geometric soft theorem and prove it.  Next, in \Sec{sec:examples} we verify the validity of this soft theorem for salient examples, including theories exhibiting general potential, two-derivative, and higher-derivative interactions.  For the case of theories with symmetry, we dedicate a longer discussion in \Sec{sec:coset}, {\it e.g.}~describing how the symmetries of the Lagrangian are described by Killing vectors whose Lie derivatives annihilate all scattering amplitudes. In addition, we explore the soft limits of Nambu-Goldstone bosons (NGBs) and explain how the Adler zero and soft dilaton theorem are simple corollaries of our geometric soft theorem.
Finally, in \Sec{sec:multisoft} and \Sec{sec:recursion}, we derive multiple-soft theorems and on-shell recursion relations.

\section{Geometry of Scalars}

\label{sec:geometry}

To begin, let us present a brief primer on the underlying geometric structure of scalar field theories.   The bulk of our discussion will recapitulate known facts, though we will describe some new results, in particular relating to the geometry of fields versus physical scattering states.  
The upshot of this exposition is that on-shell scattering amplitudes can be expressed purely in terms of the natural geometric invariants appropriate to scalar field space.

\subsection{Lagrangian and Field Basis}

Any theory of scalars $\Phi^{\cI}$ is described by a Lagrangian of the form
\eq{
   \Lagr = \tfrac{1}{2} g_{\cI\cJ}(\Phi)  \partial_\mu \Phi^\cI \partial^\mu \Phi^\cJ - V(\Phi) 
   + \tfrac{1}{4} \lambda_{\cI\cJ\cK\cL}(\Phi)  \partial_\mu \Phi^\cI \partial^\mu \Phi^\cJ \partial_\nu \Phi^\cK \partial^\nu \Phi^\cL + \cdots,
}{L_general}
where $g_{\cI\cJ}(\Phi)$ is a symmetric, non-degenerate matrix function of the scalars encoding all possible two-derivative couplings.  Furthermore, $V(\Phi)$ is a general potential while $\lambda_{\cI\cJ\cK\cL}(\Phi)$ and 
the terms in ellipses denote general higher-derivative operators. For convenience we will refer to the indices $\cI,\cJ, \cK$, etc.~as flavors, even in the absence of any underlying symmetry.

The scalar fields are maps from points in spacetime to points in a field space (or target space) describing a manifold $\cal M$.  Under field redefinitions (or coordinate transformations in field space) the scalar fields transform as
\eq{
\Phi^\cI \quad \to \quad \Phi'^\cI \, ,
}{}
where $\Phi'$ is a function of $\Phi$, while their derivatives transform as tensors via
\eq{
\partial_\mu \Phi^\cI  \quad \to \quad \frac{\partial \Phi'^\cI}{\partial \Phi^\cJ} \partial_\mu \Phi^\cJ\,.
}{}
As is well-known, $g_{\cI\cJ}(\Phi)$ can be identified as a metric on $\cal M$ that transforms as a tensor under field redefinitions,
\eq{
g_{\cI\cJ}(\Phi) \quad \to\quad  \frac{\partial \Phi^\cK}{\partial \Phi'^\cI} \frac{\partial \Phi^\cL}{\partial \Phi'^\cJ} g_{\cK\cL}(\Phi')\,.
}{}
The metric in turn defines a corresponding covariant derivative (or Levi-Civita connection), $\nabla_\cI$, together with Christoffel symbols, $\Gamma_{\cI\cJ\cK}(\Phi)$, and Riemann curvature, $R_{\cI\cJ\cK\cL}(\Phi)$. 

Under what conditions is the coupling for a given Lagrangian interaction actually a tensor in field space?  Obviously, the potential is a scalar
\eq{
 V(\Phi) &\quad \to \quad V(\Phi') \,  ,
}{}
while the higher-derivative coupling from \Eq{L_general} is a tensor,
\eq{ 
\lambda_{\cI\cJ\cK\cL}(\Phi)  &\quad \to \quad \frac{\partial \Phi^\cP}{\partial \Phi'^\cI} \frac{\partial \Phi^\cQ}{\partial \Phi'^\cJ} \frac{\partial \Phi^\cR}{\partial \Phi'^\cK} \frac{\partial \Phi^\cS}{\partial \Phi'^\cL} \lambda_{\cP\cQ\cR\cS}(\Phi') \, .
}{} 
In general, any Lagrangian interaction that depends solely on $\Phi^I$ and its first derivative $\partial_\mu \Phi^I$ will appear with a coupling that is a tensor.  This is true because $\partial_\mu \Phi^I$ is already itself a tensor.  There can, however, exist interactions with {\it more} than one derivative per field, {\it e.g.}~involving the non-tensorial object $\partial_\mu \partial_\nu \Phi^I$.  The couplings associated with these interactions are not tensors.  As is well known, on-shell scattering amplitudes are invariant under changes of field basis, so these non-tensorial couplings will always enter in combinations that behave precisely as tensors.

\subsection{Vacuum}

Of course, in order to compute physical quantities it will be necessary to choose a vacuum at which the fields reside.  To this end we define
\eq{
\Phi^I = v^I + \phi^I \, ,
}{}
where $\phi^I$ describe dynamical fluctuations about constant VEVs of the scalar fields, $v^I$.  All of the dynamics of $\phi^I$ are dictated by the Lagrangian couplings and their derivatives evaluated at the VEV.  So for example, their dispersion relations are controlled by $g_{IJ}(v)$ and $\partial_I \partial_J V(v)$---so crucially we have {\it not} assumed that the scalar fields are canonically normalized or that the mass matrix is diagonalized.  We will return to this point later on.  Meanwhile, the cubic interactions arise from $\partial_I g_{JK}(v)$ and $\partial_I \partial_J \partial_K V(v)$, and so on and so forth.  Since the vacuum is assumed to be stable, the tadpole $\partial_I V(v)$ is zero.  

In general, off-shell quantities such as the interaction vertices or correlators are {\it not} covariant under field redefinitions.  However, as noted previously, observables such as the on-shell scattering amplitudes are indeed covariant. 

For notational convenience, we will sometimes denote symmetrized covariant derivatives of the potential by 
\eq{
V_{I_1 \cdots I_n}(v) =  \nabla_{(I_1} \cdots \nabla_{I_n)} V(v)\, ,
}{}
where the right-hand size is defined to include a $1/n!$ symmetry factor.  We emphasize that the derivatives here are taken with respect to the VEVs rather than the dynamical fields.  Hereafter this will be assumed throughout, unless otherwise stated.     Since $V_I(v)=0$ for a stable vacuum, we also find that $V_{IJ}(v) =  \partial_\cI\partial_\cJ V(v)$ and $V_{IJK}(v)  = \nabla_I V_{JK}(v)$.

\subsection{Fields vs.~States}

The scalar fields $\phi^I$ we have discussed thus far are {\it flavor eigenstates}\footnote{Strictly speaking, this is a slight abuse of nomenclature since these states are not even canonically normalized on account of $g_{IJ}(v)$ being completely arbitrary.}.  This basis is particularly useful for exhibiting the underlying geometry of field space.  On the other hand, we know that physical particles that undergo scattering are actually {\it mass eigenstates}.    Naively, it might seem a somewhat formal exercise to construct the  explicit mapping between these bases.  However, this is actually crucial for any analysis of soft limits in amplitudes.  The reason for this is that the soft limit requires the identification of a massless particle, which in turn requires the existence of a well-defined notion of mass.  In other words, it is not actually possible to take the soft limit of a general flavor eigenstate.

To construct the relationship between flavor and mass eigenstates we have to {\it canonically normalize and diagonalize} the linearized equation of motion for the scalar fields,
\eq{
\left( g_{\cI \cJ}(v) \Box + V_{\cI \cJ}(v) \right) \phi^{\cI}(x) =0 \,.
}{}
Amusingly, this is achieved with the aid of another well-known geometric object: the local orthonormal basis.  In particular, we introduce a tetrad $e_i{}^I(v)$ which flattens the metric via
\eq{
	g_{\cI\cJ}(v) e_{\fI}{}^{\cI}(v) e_{\fJ}{}^{\cJ}(v) = \delta_{\fI\fJ}\,  .
}{eq:vielbein}
Hereafter, the indices $\fI,\fJ,\fK$, etc.~will denote tetrad indices---or equivalently, mass eigenstate indices---that are raised and lowered using $\delta^{\fI\fJ}$ and $\delta_{\fI\fJ}$, respectively.  The inverse tetrad is related to the tetrad via $
e^\fI{}_\cI(v) =e_{\fJ}{}^{\cJ}(v)  \delta^{\fI\fJ} g_{\cJ\cI}(v)$.
\Eq{eq:vielbein} shows that the tetrads canonically normalize the kinetic terms of the scalar fields.  Meanwhile, in the tetrad basis the mass matrix is
\eq{
V_{\cI \cJ}(v)  e_{\fI}{}^{\cI}(v) e_{\fJ}{}^{\cJ}(v) = V_{\fI \fJ}(v) = m^2_\fI(v) \delta_{\fI\fJ} \,,
}{}
where in the last equality we have made the additional assumption that the tetrad also diagonalizes the mass matrix.  This is always possible by composing an arbitrary tetrad by a suitable orthogonal rotation. 

The mass eigenstates $|p^\fI\rangle $ which describe scattering particles are labelled by a momentum $p$ and tetrad index $\fI$.  The overlap of this mass eigenstate with a flavor eigenstate field is proportional to the tetrad,
\eq{
\langle p^\fI |\phi^\cJ(x)  | 0\rangle =  e^{\fI\cJ}(v) e^{ip \cdot x}\,,
}{}
where the on-shell condition is $p^2 =m^2_\fI(v)$.  

Last but not least, let us define the $n$-particle scattering amplitude,
\eq{
\Amp{n}^{i_1 \cdots i_n}(p_1, \cdots , {p_n}) \delta^D(p_1 +\cdots + p_n)  &= \langle p_1^{i_1} \cdots p_n^{i_n} | 0\rangle \, ,
}{}
which is obtained via LSZ reduction of the $n$-particle correlator,
\eq{
\langle p_1^{i_1} \cdots p_n^{i_n} | 0\rangle
 = (-i)^{n+1}  \left[ \prod_{a=1}^n  \lim_{p_a^2\rightarrow m_{\fI_a}^2} 
	 (p_a^2- m_{\fI_a}^2) e^{i_a}{}_{I_a}
\right]
\langle T \phi^{I_1}(p_1)   \cdots \phi^{I_n}(p_n) \rangle  \, .
}{eq:LSZ}
This perspective offers a new geometric interpretation for the wavefunction normalization factors in the LSZ reduction formula: they are tetrad factors.

Since our analysis leans heavily on the geometry of field space, it will sometimes be more convenient to transform amplitudes from mass eigenstate to flavor eigenstate,
\eq{
\Ampu{n}_{I_1 \cdots I_n}(p_1, \cdots , {p_n})  = \left[ \prod_{a=1}^n e_{i_a I_a} \right] \Amp{n}^{i_1 \cdots i_n}(p_1, \cdots , {p_n}) \, ,
}{eq:A_tetrad}
and vice versa.
It should be obvious from \Eq{eq:A_tetrad} and its inverse that it is mechanically trivial to toggle between field space and tetrad indices, or equivalently, between flavor and mass eigenstate.    Indeed, for any expression written in terms of geometric invariants, this operation is nothing more than a lexographical relabelling of indices. 

Let us comment briefly on our choice of basis for explicit computations.  Of course, since the Lagrangian is composed of objects with field-space indices, the same will be true of the Feynman rules derived from this Lagrangian.  Thus, for practical calculations it is simplest to use those Feynman rules to directly compute the amplitude in flavor eigenstate, $\Ampu{n}_{I_1 \cdots I_n}$, and in the end to transform to the amplitude in mass eigenstate, $ \Amp{n}^{i_1 \cdots i_n}$.  The latter is the physical quantity that is subject to our soft theorem.

\section{Geometric Soft Theorem}

\label{sec:theorem}

Armed with a framework describing the underlying geometry of scalar fields, we are now ready to introduce the main result of this paper: a geometric soft theorem valid for any massless state in a general theory of scalars.

\subsection{Statement of Theorem}

Our claim is that the soft limit of the tree-level $(n+1)$-particle amplitude is universally related to the tree-level $n$-particle amplitude via\footnote{We believe that this soft theorem actually applies non-perturbatively in the case of vanishing potential.}
\eq{
  \lim_{q\rightarrow 0}\Amp{n+1}^{i_1\cdots i_n i} 
  =& \phantom{{}+{}} \nabla^i \Amp{n}^{i_1\cdots i_n} 
 +  \sum_{a=1}^n 
  \frac{ \nabla^i V^{i_a}{}_{j_a}}{(p_a +q)^2  - m_{j_a}^2} \left(1+ q^\mu \frac{\partial}{\partial p_a^\mu} \right)
\Amp{n}^{i_1 \cdots j_a \cdots i_n} \, .
}{eq:soft_thm}
Here the soft particle is a massless scalar carrying momentum $q$ and labelled by index $i$, and we have dropped all contributions at ${\cal O}(q)$ and higher.
As we will show, this result is applicable even in the presence of massive spectator states and arbitrary higher-derivative operators.

Let us comment briefly on the technical application of this soft theorem.  First, we emphasize again that the covariant derivative $\nabla^i$ is computed with respect to the VEVs and {\it not} the  dynamical scalars, as should be obvious since the latter do not even appear in the amplitude.   Second, in order to ensure that the soft limit itself maintains total momentum conservation, we send $q\rightarrow 0$ only after eliminating the momentum of some auxiliary leg from the $(n+1)$-particle amplitude.
Consequently,  the differential operator $ q^\mu \frac{\partial}{\partial p_a^\mu}$ should also be applied with that same prescription.  A similar approach is needed in the application of subleading soft theorems in gauge theory and gravity \cite{Bern:2014vva}. We will elaborate on this point later on. 

\Eq{eq:soft_thm} is a simple geometric restatement of the usual physical intuition: a soft scalar parameterizes a shift of the VEVs.   In particular, the first term in \Eq{eq:soft_thm} describes the effect of a shift of the VEV on coupling constants and internal propagators, while the second term corresponds to the effect of this shift on external propagators.  Notably, the latter disappears when the theory does not have a potential.

\subsection{Proof of Theorem}

We are now ready to prove the soft theorem in \Eq{eq:soft_thm}. To begin, consider the Euler-Lagrange equations of motion for the fields $\phi^\cI$ that fluctuate about $v^\cI$,
\eq{
\partial_\mu \J_\cI^\mu = \frac{\delta \Lagr}{\delta \phi^\cI } \qquad \textrm{where} \qquad \J_\cI^\mu = \frac{\delta \Lagr}{\delta \partial_\mu \phi^\cI } \, .
}{}
By construction, the dependence of the Lagrangian on the scalar field and on the VEV enter identically, since $\Phi^\cI = v^\cI + \phi^\cI$.  The equations of motion thus imply that
\eq{
\partial_\mu \J_\cI^\mu = \partial_\cI \Lagr\, ,
}{eq:EOM}
where the derivative on the right-hand side is with respect to the VEV.   
We will show that the soft theorem in \Eq{eq:soft_thm} follows from evaluating \Eq{eq:EOM} between scattering states.  

To begin, let us expand the Lagrangian in \Eq{L_general} about $\Phi^\cI = v^\cI + \phi^\cI$, yielding
\eq{
\Lagr = & \phantom{{}+{}}  \tfrac{1}{2}\left(g_{\cI\cJ}(v)+ \partial_\cK g_{\cI\cJ}(v) \phi^\cK +\cdots \right) \partial_\mu \phi^\cI \partial^\mu \phi^\cJ \\
& - \left(V(v)+\partial_\cI V(v) \phi^\cI + \tfrac{1}{2} \partial_\cI \partial_\cJ V(v) \phi^\cI \phi^\cJ +\cdots \right) + \cdots \, ,
}{L_general_exp}
where the ellipses denote terms higher order in fields or higher order in derivatives which will not affect our analysis.
From \Eq{L_general_exp} we compute the left- and right-hand sides of \Eq{eq:EOM}, 
\eq{
\partial_\mu \J^\mu_\cI &= \Box \phi_\cI + \partial_\cK g_{\cI\cJ} (v)\partial_\mu ( \phi^\cK \partial^\mu \phi^\cJ  ) + \cdots \,, \\
\partial_I \Lagr &= \tfrac{1}{2} \partial_\cI g_{\cJ\cK}(v) \partial_\mu \phi^\cJ \partial^\mu \phi^\cK - \partial_\cI \partial_\cJ V(v) \phi^\cJ - \tfrac{1}{2} \partial_\cI \partial_\cJ \partial_\cK V(v) \phi^\cJ \phi^\cK + \cdots \,.
}{ops}
We will be interested in the overlap of the above operators with $n$-particle scattering states.  These operators carry a free index $I$ which we identify with that of the soft particle.  Furthermore, the momentum $q$ entering through each operator corresponds to that of the soft particle.  So we assume that $q^2=0$ and consider the $q\rightarrow 0$ limit.  For convenience, let us define the $n$-particle overlap with an operator $\cal O$ to be
\eq{
\langle  {\cal O} \rangle \delta^D (p_1 +\cdots + p_n) = \lim_{q\rightarrow 0} \int d^D x e^{iqx} e_{\fI_1 \cI_1} \cdots e_{\fI_n \cI_n} \langle p_1^{\fI_1} ,\cdots p_n^{\fI_n} | {\cal O}(x) |0\rangle \, .
}{}
Now taking the $n$-particle overlap of the terms in the first line of \Eq{ops}, we obtain
\eq{
\langle \Box \phi_\cI\rangle &= \lim_{q\rightarrow 0}
\Ampu{n+1}_{\cI_1 \cdots \cI_n \cI}  \,,
}{}
which implements LSZ reduction on the soft particle to give the on-shell $(n+1)$-particle amplitude.  Meanwhile, we also find that
\eq{
 \langle \partial_\cK g_{\cI\cJ} \partial_\mu (  \partial^\mu \phi^\cJ \phi^\cK)\rangle &= \tfrac{1}{2} \langle \partial_\cK g_{\cI\cJ} \Box (\phi^\cJ \phi^\cK)\rangle -  \tfrac{1}{2} \langle \partial_\cK g_{\cI\cJ} \partial_\mu (  \phi^\cJ \overset{\leftrightarrow}{\partial^\mu} \phi^\cK)\rangle \\
&= -  \tfrac{1}{2} \langle \partial_\cK g_{\cI\cJ}  \partial_\mu (  \phi^\cJ \overset{\leftrightarrow}{\partial^\mu} \phi^\cK)\rangle_{\rm ext}
\, ,
}{}
dropping the term involving $\Box(\phi^\cI \phi^\cJ)$ since it is proportional to $q^2=0$.  Moreover, since the term $\partial_\mu (  \phi^\cJ \overset{\leftrightarrow}{\partial^\mu} \phi^\cK)$ scales manifestly as ${\cal O}(q)$, we know that as $q\rightarrow 0$ it can only contribute when inserted on an external propagator of a lower-point amplitude.  Only in this case can a soft pole going as ${\cal O}(1/q)$ arise from the external propagator, thus yielding a non-vanishing contribution.   We denote contributions of this type with a corresponding subscript, so the operator insertion $\langle {\cal O} \rangle_{\rm ext}$ corresponds to the diagram in Fig.~\ref{fig:O_ext}.  Putting all of these pieces together we obtain
\eq{
\langle \partial_\mu \J^\mu_\cI \rangle &= \lim_{q\rightarrow 0} 
\Ampu{n+1}_{\cI_1 \cdots \cI_n \cI} 
-  \tfrac{1}{2} \langle \partial_\cK g_{\cI\cJ}  \partial_\mu (  \phi^\cJ \overset{\leftrightarrow}{\partial^\mu} \phi^\cK)\rangle_{\rm ext}\, ,
}{EOM_LHS}
for the $n$-particle overlap of the first line of \Eq{ops}.

\begin{figure}
    \centering

    \begin{tikzpicture}
            \coordinate (d) at (-3.75, 0);
            \coordinate (c) at (-3, 0);	
	    	\coordinate (a) at (-2, 1);
		        \node[left]  at (d) {$\langle {\cal O} \rangle_{\rm ext} \quad = \quad \mathlarger{ \sum_{a=1}^n}$};
	        \node[left]  at (c) {$a$};
	    	
	    	\coordinate (f) at (1, 1);
	    	\coordinate (l) at (1, -1);
	    	
	    	\coordinate (m4) at (-2,0);
	    	
	    	\coordinate (mn) at (-0.25, 0);
	    
	    	\coordinate (d1) at (1.00, 0);
	    	\coordinate (d2) at (0.85, 0.5);
	    	\coordinate (d3) at (0.85, -0.5);
	    	
	    	\draw [hard] (c) -- (m4);
	    	\draw [hard] (mn) -- (m4);
	    	
	    	\draw [hard] (f) -- (mn);
	        \draw [hard] (l) -- (mn);
	    	
            \draw[fill=lightgray, opacity=1] (mn) circle (0.65);
            	\node at (mn) {$\Amp{n}$};
            
            \draw[fill=white, opacity=1] (m4) circle (0.2);
            \node[below]  at (-2,-0.2) {${\cal O}(q) $};
	    	
            \draw[fill=black, opacity=1] (d1) circle (0.02);
            \draw[fill=black, opacity=1] (d2) circle (0.02);
            \draw[fill=black, opacity=1] (d3) circle (0.02);
\end{tikzpicture}
    \caption{Diagrams computing $\langle {\cal O}\rangle_{\rm ext}$, which sums over the insertion of an operator ${\cal O}$ on each external leg $a$ of the $n$-particle amplitude.}
    \label{fig:O_ext}
\end{figure}

Let us now compute the $n$-particle overlap of the second line of \Eq{ops}.  It will be crucial to distinguish between terms in the Lagrangian which are quadratic in fields, corresponding to kinetic and mass terms, versus those which are cubic order or higher, corresponding to interaction vertices.  The latter contribute to $\langle \partial_\cI \Lagr \rangle$ through $n$-particle diagrams in which the coupling in each interaction vertex is replaced with the {\it derivative} of that coupling with respect to the VEV.   Meanwhile, the former contribute to $\langle  \partial_\cI  \Lagr \rangle$ through $n$-particle diagrams in which a propagator is replaced with the {\it derivative} of that propagator---or more precisely, the derivative of the kinetic and mass parameters in that propagator---with respect to the VEV.   Thus we learn that 
\eq{
\langle  \partial_\cI  \Lagr  \rangle &= \partial_\cI 
\Ampu{n}_{\cI_1 \cdots \cI_n } 
+  \tfrac{1}{2} \langle \partial_\cI g_{\cJ\cK} \partial_\mu \phi^\cJ \partial^\mu \phi^\cK -  \partial_\cI \partial_\cJ \partial_\cK V \phi^\cJ \phi^\cK \rangle_{\rm ext}\,,
}{EOM_RHS}
where we have used that  $\langle \partial_\cI \partial_\cJ V(v) \phi^\cJ\rangle$ will eventually vanish when transforming to mass eigenstate basis since the soft particle is assumed to be massless.  Indeed, if the soft particle was massive then a shift of the VEV would induce a tadpole, which would be inconsistent.
The first term in this expression arises from insertions of $\partial_\cI \Lagr$ into interaction vertices or internal propagators.  Those graphical elements are the only objects that appear in the scattering amplitude, since external propagators are amputated.  Meanwhile, the second term appears when $\partial_\cI \Lagr$ is inserted on an external propagator, corresponding to the deformation of the LSZ operation from a shift of the VEV.

Equating \Eq{EOM_LHS} and \Eq{EOM_RHS} by the equations of motion and rearranging terms, we obtain 
\eq{
 \lim_{q\rightarrow 0} 
 \Ampu{n+1}_{\cI_1 \cdots \cI_n \cI} 
 & = 
 \partial_\cI \Ampu{n}_{\cI_1 \cdots \cI_n } 
+  \tfrac{1}{2} \langle  \partial_\cK g_{\cI\cJ}  \partial_\mu (  \phi^\cJ \overset{\leftrightarrow}{\partial^\mu} \phi^\cK) + \partial_\cI g_{\cJ\cK}\partial_\mu \phi^\cJ \partial^\mu \phi^\cK -  \partial_\cI \partial_\cJ \partial_\cK V \phi^\cJ \phi^\cK \rangle_{\rm ext} \\
& = 
\partial_\cI \Ampu{n}_{\cI_1 \cdots \cI_n } 
-   \langle   \Gamma_{\cK \cI\cJ} \phi^\cJ \Box  \phi^\cK  +  \tfrac{1}{2} \partial_\cI  V_{\cJ \cK} \phi^\cJ \phi^\cK \rangle_{\rm ext} \,,
}{}
where we have used that the Christoffel symbol is $ \Gamma^I{}_{JK} = \tfrac12 g^{IL} (\partial_J g_{LK} + \partial_Kg_{LJ}   - \partial_Lg_{JK})$.    Note that the operator inside the brackets is the cubic vertex for the scalar fields.
By rewriting $\Box$ as the linearized wave equation plus a difference term, we find that the latter contribution exactly combines with $ \partial_\cI  V_{\cJ \cK} $ to make it covariant, so
\eq{
 \lim_{q\rightarrow 0} 
  \Ampu{n+1}_{\cI_1 \cdots \cI_n \cI} 
  & = 
\partial_\cI \Ampu{n}_{\cI_1 \cdots \cI_n } 
-   \langle   \Gamma_{\cK \cI\cJ} \phi^\cJ (\Box  \phi^\cK + V^{\cK}{}_{ \cL} \phi^\cL)  +  \tfrac{1}{2}  \nabla_\cI V_{ \cJ \cK} \phi^\cJ \phi^\cK \rangle_{\rm ext} \,.
 }{}
As noted earlier, the remaining operator insertions should be evaluated on an external propagator connecting to an $n$-particle subamplitude.   By construction, we have arranged so that one of the terms will immediately pinch a neighboring propagator, so
\eq{
 -   \langle   \Gamma_{\cK \cI\cJ} \phi^\cJ (\Box  \phi^\cK + V^{\cK}_{ \cL} \phi^\cL)  \rangle_{\rm ext} = -\sum_{a=1}^n  \Gamma^{\cJ_a}{}_{  \cI_a \cI}\Ampu{n}_{\cI_1 \cdots \cJ_a \cdots \cI_n} \, . 
}{}
This then implies the following relation between the soft limit of the $(n+1)$-particle amplitude and the $n$-particle amplitude.
\eq{
 \lim_{q\rightarrow 0} 
\Ampu{n+1}_{\cI_1 \cdots \cI_n \cI} 
  & = 
\nabla_\cI \Ampu{n}_{\cI_1 \cdots \cI_n } -   \tfrac{1}{2} \langle   \nabla_\cI V_{ \cJ \cK} \phi^\cJ \phi^\cK \rangle_{\rm ext} \,.
}{eq:soft_thm_part1}
The last term corresponds to all contributions to the $(n+1)$-particle amplitude which arise from the cubic potential vertex attached to an $n$-particle subamplitude.  Notably, the latter is actually an {\it off-shell} object, since one of the external legs has accumulated the soft momentum $q$.  The potential contribution can be evaluated explicitly, and is 
\begin{align}
 - \tfrac{1}{2}   \langle   V_{\cI \cJ \cK} \phi^\cJ \phi^\cK \rangle_{\rm ext} &= \sum_{a=1}^n  \nabla_I V_{I_a  }{}^{J_a}
\Delta(p_a+q)_{J_a}{}^{K_a}
\Ampu{n}_{\cI_1 \cdots \cK_a \cdots \cI_n}(p_1,\cdots, p_a+q, \cdots, p_n) \label {eq:soft_thm_part2}\\
&= \sum_{a=1}^n  \nabla_I V_{I_a  }{}^{J_a}
\Delta(p_a+q)_{J_a}{}^{K_a}\left(1+ q^\mu \frac{\partial}{\partial p_a^\mu} \right)
\Ampu{n}_{\cI_1 \cdots \cK_a \cdots \cI_n}(p_1,\cdots, p_a, \cdots, p_n) \, ,\nonumber
\end{align}
where $\Delta(p)_I{}^J = (\delta^I{}_J p^2 - V^I{}_J)^{-1}$ is the propagator in the flavor eigenstate basis.
In the first line, we have technically taken an off-shell continuation of the $n$-particle amplitude since the external leg carrying the operator insertion carries momentum $p_a +q$ accrued from the momentum $q$ of the soft particle.   Since this object is off-shell, one might rightly worry that this expression is not invariant under changes of field basis.  As we will see in the next section, however, these off-shell contributions magically cancel amongst each other in the full soft theorem.  Moreover, in the second line above we have simply rewritten the shifted momentum as an operator $1+ q^\mu \partial /\partial p_a^\mu$ acting on the on-shell $n$-particle amplitude.   Combining \Eq{eq:soft_thm_part1} and \Eq{eq:soft_thm_part2} and trivially mapping to tetrad basis, we obtain the claimed geometric soft theorem in \Eq{eq:soft_thm}.

\subsection{Field Basis Independence}

It is not a priori obvious why the soft theorem in \Eq{eq:soft_thm} is a sensible on-shell operation.  The left-hand side is the limit of an on-shell quantity, but the right-hand side is the {\it differential} of an on-shell quantity.  If the derivative in question does not preserve the on-shell conditions then one could rightly worry that the resulting expression is inherently off-shell and thus dependent on choice of field basis.  Here we address those concerns and show that they are not a problem.

First, we study the question of total momentum conservation.  Consider the addition of ``zero'' to the $n$-particle amplitude by adding
\eq{
\delta \Amp{n}^{\fI_1\cdots \fI_n} = (p_1 +\cdots + p_n)^\mu {\cal O}^{\fI_1\cdots \fI_n}_\mu \, ,
}{eq:zero1}
which is simply the total momentum contracted with an arbitrary tensor.  By construction, $\delta \Amp{n}^{\fI_1\cdots \fI_n}=0$ is zero on-shell, so we would be permitted to include or discard it at our convenience.  However, let us instead consider the action of the soft theorem in \Eq{eq:soft_thm} on this contribution.  Obviously, $\nabla^\fI  \delta \Amp{n}^{\fI_1\cdots \fI_n}=0$ since $\nabla^\fI$ commutes with total momentum.  On the other hand, it is evident that $q^\mu \frac{\partial}{\partial p_a^\mu}  \delta \Amp{n}^{\fI_1\cdots \fI_n}\neq0$.  Thus,  the soft theorem naively does not preserve total momentum conservation.  

The resolution to this issue is that the soft limit itself requires a prescription in order to preserve momentum conservation.  This is because sending $q\rightarrow 0$ with all other momenta fixed does not preserve total momentum conservation. Consequently, the soft limit must be defined with the prescription that the momentum of some auxiliary particle---distinct from the soft particle---has been eliminated from the $(n+1)$-particle amplitude to begin with.  The same prescription should then be applied to the $n$-particle amplitude, in which case \Eq{eq:zero1} is algebraically zero.

Second, let us consider the on-shell conditions.  Again we add ``zero'' to the $n$-particle amplitude, this time via
\eq{
\delta \Amp{n}^{\fI_1\cdots \fI_n} = \Delta^{-1}(p_a)^{\fI_a}{}_{\fJ_a} {\cal O}^{\fI_1\cdots \fJ_a \cdots \fI_n} = (\delta^{\fI_a}{}_{\fJ_a} p_a^2 - V^{\fI_a}{}_{\fJ_a}) {\cal O}^{\fI_1\cdots \fJ_a \cdots \fI_n} \, ,
}{eq:zero2}
which is the on-shell condition for particle $a$ contracted with an arbitrary tensor.  Plugging $\delta \Amp{n}^{\fI_1\cdots \fI_n}$ into the soft theorem in \Eq{eq:soft_thm}, we find that the first term is
\eq{
-\nabla^\fI V^{\fI_a}{}_{\fJ_a}  {\cal O}^{\fI_1\cdots \fJ_a \cdots \fI_n} \, ,
}{}
while the second term evaluates to
\eq{
\nabla^\fI V^{\fI_a  }{}_{\fJ_a}
\Delta(p_a+q)^{\fJ_a}{}_{\fK_a} \Delta^{-1}(p_a+q)^{\fK_a}{}_{\fL_a}  {\cal O}^{\fI_1\cdots \fL_a \cdots \fI_n} = \nabla^\fI V^{\fI_a}{}_{\fJ_a}  {\cal O}^{\fI_1\cdots \fJ_a \cdots \fI_n} \, ,
}{}
where we have ignored all terms in which the differentials act on factors other than the on-shell condition.  Remarkably, we see that the first and second terms exactly cancel, so $\delta \Amp{n}^{\fI_1\cdots \fI_n} $ has no effect on the soft theorem.

In general, one might worry that a factor of the on-shell condition can appear in the $n$-particle amplitude in more complicated ways than in \Eq{eq:zero2}.   For example, the on-shell condition might appear quadratically or higher.  Fortunately, this scenario is not an issue because the differential operators will always leave a residual factor that is linear or higher in the on-shell condition, which will in turn vanish.  Another possibility is that the on-shell condition enters deep inside the $n$-particle amplitude, for instance through an interaction vertex.  However, even in this case, any factor of $p_a^2$ which appears can be reassociated with the $\delta^{\fI_a}{}_{\fJ_a}$ that necessarily sits out in front of the amplitude, in which case \Eq{eq:zero2} applies once again.

To summarize, we have shown that the soft theorem in \Eq{eq:soft_thm} preserves on-shell kinematics, and is thus a sensible operation.  As a corollary, all quantities on the left- and right-hand sides of this formula will appear in combinations which are invariant under changes of field basis.

\section{Examples}

\label{sec:examples}

In this section we apply the geometric soft theorem to some concrete examples.  Of course, one can study a theory of maximal generality such as the one defined in \Eq{L_general}, which includes arbitrary potential terms as well as two- and higher-derivative interactions.  In fact, we have verified via explicit calculation that our soft theorem is valid up to seven-particle scattering in this general theory.  Unsurprisingly, the resulting amplitudes expressions are quite complicated and not particularly illuminating.  So it will be more instructive to instead consider the soft theorem in some explicit examples in which only a subset of the couplings is active.

\subsection{General Potential}

To start, consider a theory with a flat field-space manifold  but with an arbitrary potential,
\eq{
	L &= \tfrac{1}{2} \delta_{\cI\cJ}  \partial_{\mu} \Phi^{\cI} \partial^{\mu} \Phi^{\cJ}  - V(\Phi) \, .
}{}
Since the field-space metric is flat, the tetrad is trivial and the covariant derivative, $\nabla_I$, and partial derivative, $\partial_I$, are interchangeable.
A straightforward computation yields the three-, four-, and five-particle  scattering amplitudes,
\eq{
\Amp{3}^{\fI_1 \fI_2 \fI_3} =& - V^{\fI_1 \fI_2 \fI_3} \,,\\
\Amp{4}^{\fI_1 \fI_2 \fI_3 \fI_4} =& - V^{\fI_1 \fI_2 \fI_3 \fI_4} - \sum_\fJ \left ( \frac{ V^{ \fI_1 \fI_2 \fJ }V_{\fJ}{}^{\fI_3 \fI_4} }{s_{12} - m_\fJ^2}  +\frac{ V^{ \fI_2 \fI_3  \fJ}V_{\fJ}{}^{\fI_1 \fI_4} }{s_{23} - m_\fJ^2}  + \frac{ V^{ \fI_1 \fI_3 \fJ}V_{\fJ}{}^{\fI_2 \fI_4} }{s_{13} - m_\fJ^2}   \right)  \,, \\
\Amp{5}^{\fI_1 \fI_2 \fI_3 \fI_4 \fI_5} =& - V^{\fI_1 \fI_2 \fI_3 \fI_4 \fI_5} -
\sum_{\fJ} \left(\frac{V^{\fI_{1} \fI_{2} \fJ} V_{\fJ}{}^{\fI_3 \fI_4 \fI_5}}{s_{12} - m^{2}_{\fJ}} + {\rm perm.}\right) \\ &- \sum_{\fJ,\fK} \left(\frac{V^{\fI_1 \fI_2 \fJ} V_{\fJ}{}^{ \fI_3 \fK} V_{\fK}{}^{\fI_4 \fI_5}}{(s_{12} - m^{2}_{\fJ})(s_{45} - m^{2}_{\fK})} + {\rm perm.} \right) \, ,
}{}
where here and throughout we define the Mandelstam invariants to be $s_{ij} = (p_i + p_j)^2$ and $s_{ijk} = (p_i + p_j + p_k)^2$. 

We can now verify the validity of the soft theorem in \Eq{eq:soft_thm}.  For example,  consider the $p_4\rightarrow 0$ limit of the four-particle amplitude.  The four-particle contact contribution $- V^{\fI_1 \fI_2 \fI_3 \fI_4}$ maps correctly onto $-\partial^{\fI_4} V^{\fI_1 \fI_2 \fI_3}$.  Furthermore,  the four-particle factorization contributions map correctly onto the terms in \Eq{eq:soft_thm} involving the cubic potential terms.  As noted previously, these terms have the interpretation as the variation of the external propagators with respect to the VEV.    The soft theorem similarly holds for the $p_5 \rightarrow 0$ soft limit of the five-particle amplitude, and so on and so forth.

The above analysis incorporates all possible potential operators.  It is of course natural to ask what happens if we restrict to a subset of interactions, for example as would arise in $ \Phi^4$ theory.  However, we should emphasize here that our geometric soft theorem requires the inclusion of {\it all interactions} generated by an infinitesimal shift of the VEV.  This is necessary because the covariant derivative effectively probes a small region in field space about the VEV.  So to apply our geometric soft theorem to $\Phi^4$ theory one would need to compute amplitudes in a neighborhood of the original theory in field space, which in effect requires including $\Phi^3$ operators as well.

\subsection{Two-Derivative Interactions}

\label{subsec:twoder}

Next, let us examine a theory of massless scalars with general two-derivative interactions and no potential, also known as the non-linear sigma model (NLSM).  The Lagrangian is
\eq{
	L &= \tfrac{1}{2} g_{\cI\cJ}(\Phi) \partial_{\mu} \Phi^{\cI} \partial^{\mu} \Phi^{\cJ}
	  = \tfrac{1}{2} (\delta_{\cI\cJ} - \tfrac{1}{3} R_{\cI\cK\cJ\cL} \phi^\cK\phi^\cL - \tfrac{1}{6} \nabla_\cK R_{\cI\cL\cJ\cM}\phi^\cK \phi^\cL \phi^\cM +\cdots)  \partial_{\mu} \phi^{\cI} \partial^{\mu} \phi^{\cJ} \, ,
}{L_GNLSM}
where in the second equation we have expanded about the VEV using the analog of Riemann normal coordinates.  This choice of field basis makes the dependence on geometric quantities manifest.  Note that the three-particle vertex is identically zero, since the Christoffel connection evaluated at the vacuum is zero in Riemann normal coordinates.  This fact can also be understood purely from the perspective of amplitudes: the on-shell three-particle amplitude in a derivatively coupled theory of scalars is always zero, so the corresponding Lagrangian term can be eliminated by a field redefinition. 

It is a simple but tedious exercise to compute the three-, four-, five-, and six-particle scattering amplitudes, 
\eq{
	\Amp{3}^{\fI_1 \fI_2 \fI_3}  \phantom{{}+{}} =&  \phantom{{}+{}}  0 , \\
	\Amp{4}^{\fI_1 \fI_2 \fI_3 \fI_4}  \phantom{{}+{}} =& \phantom{{}+{}} R^{\fI_1 \fI_3 \fI_2 \fI_4 } s_{34} + R^{\fI_1 \fI_2 \fI_3 \fI_4} s_{24} , \\
	\Amp{5}^{\fI_1 \fI_2 \fI_3 \fI_4 \fI_5}  \phantom{{}+{}} =&  \phantom{{}+{}}  \nabla^{\fI_3} R^{\fI_1 \fI_4 \fI_2 \fI_5} s_{45} + \nabla^{\fI_4} R^{\fI_1 \fI_3 \fI_2 \fI_5} s_{35} + \nabla^{\fI_4} R^{\fI_1 \fI_2 \fI_3 \fI_5} s_{25} \\& + \nabla^{\fI_5} R^{\fI_1 \fI_3 \fI_2 \fI_4} s_{34} + \nabla^{\fI_5} R^{\fI_1 \fI_2 \fI_3 \fI_4} (s_{24} + s_{45}) , \\
\Amp{6}^{\fI_1 \fI_2 \fI_3 \fI_4 \fI_5 \fI_6}  \phantom{{}+{}} =& - \tfrac{1}{72} \left(R^{\fI_1 \fI_3 \fI_2 \fJ} s_{12} + R^{\fI_1 \fI_2 \fI_3 \fJ} s_{13}  \right) \frac{1}{s_{123}}  \left( R_{\fJ}^{\;\;\fI_6 \fI_5 \fI_4} s_{46} + R_{\fJ}^{\;\;\fI_5 \fI_6 \fI_4} s_{45}\right)\\&
	+ \tfrac{1}{108} \left(R^{\fI_1 \fI_3 \fI_2 \fJ} (s_{12} - \tfrac{1}{6}s_{123}) + R^{\fI_1 \fI_2 \fI_3 \fJ} (s_{13} - \tfrac{1}{6}s_{123})  \right) \left( R_{\fJ}^{\;\; \fI_6 \fI_5 \fI_4}  + R_{\fJ}^{\;\; \fI_5 \fI_6 \fI_4} \right) \\&
	+ \tfrac{1}{90} R^{\fI_1 \fI_6 \fI_5 \fJ} R_{\fJ}^{\;\; \fI_2 \fI_3 \fI_4} s_{13} + \tfrac{1}{80} \nabla^{\fI_6} \nabla^{\fI_5} R^{\fI_1 \fI_2 \fI_3 \fI_4} s_{13}
	+ {\rm perm}\, .
}{GNLSM_amps}
We are now equipped to verify the validity of the soft theorem in \Eq{eq:soft_thm} for the two-derivative theory.\footnote{Note that in all subsequent analysis it is necessary to reduce to a minimal basis of Riemann curvature tensors and their derivatives using the first and second Bianchi identities. Furthermore one must place covariant derivatives in a canonical order.}  
The $p_4 \rightarrow 0$ limit of the four-particle amplitude is zero,
\eq{
	\lim_{p_4\rightarrow 0} \Amp{4}^{\fI_1 \fI_2 \fI_3 \fI_4} &=0 = \nabla^{\fI_4} \Amp{3}^{\fI_1 \fI_2 \fI_3 } \, , 
}{}
which is consistent since the three-particle amplitude vanishes identically.  Meanwhile, taking the $p_5 \rightarrow 0$ limit of the five-particle amplitude yields
\eq{
	\lim_{p_5\rightarrow 0} \Amp{5}^{\fI_1 \fI_2 \fI_3 \fI_4 \fI_5} &= \nabla^{\fI_5} R^{\fI_1 \fI_3 \fI_2 \fI_4} s_{34} + \nabla^{\fI_5} R^{\fI_1 \fI_2 \fI_3 \fI_4} s_{24} = \nabla^{\fI_5} \Amp{4}^{\fI_1 \fI_2 \fI_3 \fI_4} \, ,
}{GNLSM_5pt}
which again accords with the soft theorem. 
Last of all, the $p_6 \rightarrow 0$ limit of the six-particle amplitude also satisfies the soft theorem in a quite non-trivial fashion, as factorizable and non-factorizable terms conspire to give the covariant derivative of the five-particle amplitude.  While the above analysis applies to any two-derivative theory, we will discuss in \Sec{sec:coset} the case where the soft limit vanishes, {\it i.e.}~there is an Adler zero.

\subsection{Higher-Derivative Interactions}

\label{subsec:lambda}

Last but not least, let us consider the two-derivative theory augmented by the leading higher-derivative interaction.   This theory is described by the Lagrangian,
\eq{
	L &= \tfrac{1}{2} g_{\cI\cJ}(\Phi) \partial_{\mu} \Phi^{\cI} \partial^{\mu} \Phi^{\cJ} +\tfrac{1}{4}\lambda_{\cI\cJ\cK\cL}(\Phi)  \partial_\mu \Phi^\cI \partial^\mu \Phi^\cJ \partial_\nu \Phi^\cK \partial^\nu \Phi^\cL \, .
}{LhigherD}
The $\lambda$-dependent contributions to three-, four-, five-, and six-particle scattering amplitudes are 
\eq{
\Amp{3,\lambda}^{\fI_1 \fI_2 \fI_3} 
=& \phantom{{}+{}} 0 \,, \\
\Amp{4,\lambda}^{\fI_1 \fI_2 \fI_3 \fI_4}  
=& \phantom{{}+{}}  \tfrac{1}{2} s_{12} s_{34} \lambda^{\fI_1 \fI_2 \fI_3 \fI_4} + \tfrac{1}{2} s_{13} s_{24} \lambda^{\fI_1 \fI_3 \fI_2 \fI_4} + \tfrac{1}{2} s_{23} s_{14} \lambda^{\fI_2 \fI_3 \fI_1 \fI_4} \,, \\
\Amp{5,\lambda}^{\fI_1 \fI_2 \fI_3 \fI_4 \fI_5} 
=&  \phantom{{}+{}}   \tfrac{1}{2} s_{12} s_{34} \nabla^{\fI_5} \lambda^{\fI_1 \fI_2 \fI_3 \fI_4} + \tfrac{1}{2} s_{13} s_{24} \nabla^{\fI_5} \lambda^{\fI_1 \fI_3 \fI_2 \fI_4} + \tfrac{1}{2} s_{23} s_{14}  \nabla^{\fI_5} \lambda^{\fI_2 \fI_3 \fI_1 \fI_4} \\
& + \tfrac{1}{2} s_{23} s_{45} \nabla^{\fI_1} \lambda^{\fI_2 \fI_3 \fI_4 \fI_5} + \tfrac{1}{2} s_{24} s_{35} \nabla^{\fI_1} \lambda^{\fI_2 \fI_4 \fI_3 \fI_5} + \tfrac{1}{2} s_{34} s_{25} \nabla^{\fI_1} \lambda^{\fI_3 \fI_4 \fI_2 \fI_5} \\
& + \tfrac{1}{2} s_{13} s_{45} \nabla^{\fI_2} \lambda^{\fI_1 \fI_3 \fI_4 \fI_5} + \tfrac{1}{2} s_{14} s_{35} \nabla^{\fI_2} \lambda^{\fI_1 \fI_4 \fI_3 \fI_5} + \tfrac{1}{2} s_{34} s_{15}  \nabla^{\fI_2} \lambda^{\fI_3 \fI_4 \fI_1 \fI_5} \\
& + \tfrac{1}{2} s_{12} s_{45} \nabla^{\fI_3} \lambda^{\fI_1 \fI_2 \fI_4 \fI_5} + \tfrac{1}{2} s_{14} s_{25} \nabla^{\fI_3} \lambda^{\fI_1 \fI_4 \fI_2 \fI_5} + \tfrac{1}{2} s_{24} s_{15}  \nabla^{\fI_3} \lambda^{\fI_2 \fI_4 \fI_1 \fI_5} \\
& + \tfrac{1}{2} s_{12} s_{35} \nabla^{\fI_4} \lambda^{\fI_1 \fI_2 \fI_3 \fI_5} + \tfrac{1}{2} s_{13} s_{25} \nabla^{\fI_4} \lambda^{\fI_1 \fI_3 \fI_2 \fI_5} + \tfrac{1}{2} s_{23} s_{15}  \nabla^{\fI_4} \lambda^{\fI_2 \fI_3 \fI_1 \fI_5} \, , \\
\Amp{6,\lambda}^{\fI_1 \fI_2 \fI_3 \fI_4 \fI_5}  
=& 
- \tfrac{1}{32} \left( s_{12} ( s_{13} + s_{23})\lambda^{\fI_1 \fI_2 \fI_3 \fJ} \right) \frac{1}{s_{123}}\left(s_{45} ( s_{46} + s_{56})\lambda_{\fJ}^{\;\; \fI_6  \fI_5  \fI_4 } \right)
\\ &
+ \tfrac{1}{24} \left( s_{12} ( s_{13} + s_{23})\lambda^{\fI_1 \fI_2 \fI_3 \fJ} \right) \frac{1}{s_{123}} \left(R_{\fJ}^{\;\; \fI_6 \fI_5 \fI_4} (s_{46} - \tfrac{1}{3}s_{456})  + R_{\fJ}^{\;\; \fI_5 \fI_6 \fI_4} (s_{45} - \tfrac{1}{3}s_{456})  \right)
\\ &
+ \tfrac{1}{48} s_{12}s_{34} \left(R^{\fI_4 \fI_6 \fI_5 \fJ} \lambda_{\fJ}^{\;\; \fI_3 \fI_2 \fI_1}  +  R^{\fI_2 \fI_6 \fI_5 \fJ} \lambda_{\fJ}^{\;\; \fI_1 \fI_4 \fI_3 }  \right)
+ \tfrac{1}{32}s_{12}s_{34} \nabla^{\fI_6} \nabla^{\fI_5} \lambda^{\fI_1 \fI_2 \fI_3 \fI_4} + {\rm perm.}
\,,
}{}
which should be added to the two-derivative amplitudes computed in \Eq{GNLSM_amps}.  In accordance with \Eq{eq:soft_thm}, the $p_{4} \rightarrow 0$ soft limit of the four-particle amplitude vanishes.   Furthermore, the $p_{5} \rightarrow 0$ soft limit for the five-particle scattering amplitude also satisfies the soft theorem. 
The soft theorem also holds for the $p_6 \rightarrow 0$ soft limit of the six-particle amplitude, with delicate cancellations between contact terms and factorizable terms.

\section{Theories with Symmetry}

\label{sec:coset}

Historically, soft theorems were discovered in the context of spontaneous symmetry breaking.  For each broken generator of the internal symmetry, a NGB emerges at low energies.  This state transforms non-linearly under the symmetry and can exhibit certain universal soft behaviors.

For these reasons we dedicate the present section solely to the application of our geometric soft theorems to theories with symmetry.  Our analysis will apply to any theory whose scalars exhibit a symmetry of linear or non-linear type.  The case of NGBs will be a subset of our results.  As we will see, our framework offers a new unified perspective on well-established phenomena such as the Adler zero and the dilaton soft theorem.

\subsection{Geometry of Symmetry}
	
Let us consider a general theory of scalars that is invariant under the following symmetry
 transformation of the fields,
\eq{
         \Phi^\cI \quad\to \quad \Phi^\cI + \Kill^\cI(\Phi)\, .
}{eq:globalsymm}
The symmetry in question may be linear or non-linear.  Invariance implies that the Lagrangian is unchanged under \Eq{eq:globalsymm}. Applying the transformation to the general Lagrangian in \Eq{L_general}, we find that the couplings are effectively shifted by
\eq{
 g_{IJ}(\Phi) &\quad\rightarrow\quad     g_{IJ}(\Phi) +     {\cal L}_\Kill  g_{IJ}(\Phi) \, ,\\
     V(\Phi) &\quad\rightarrow\quad     V(\Phi) +     {\cal L}_\Kill  V(\Phi) \,, \\
         \lambda_{IJKL}(\Phi) &\quad\rightarrow\quad     \lambda_{IJKL}(\Phi) +     {\cal L}_\Kill  \lambda_{IJKL}(\Phi) \,, 
}{}
and so on, and where ${\cal L}_K$ is the Lie derivative with respect to $\Kill^I(\Phi)$.  Since each term in \Eq{L_general} has different derivative structure, they are each separately invariant, so
\eq{
        \mathcal{L}_\Kill g_{\cI\cJ}(v) =  \mathcal{L}_\Kill V(v)  = \mathcal{L}_\Kill \lambda_{IJKL}(v) = 0\, ,
}{eq:KillV}
where we have chosen to evaluate these expressions at the VEV.  Crucially, we observe that the first condition in \Eq{eq:KillV} implies that
\eq{
     \mathcal{L}_\Kill g_{\cI\cJ}(v) = \nabla_\cI \Kill_\cJ(v) + \nabla_\cJ \Kill_\cI(v) = 0\,,
}{eq:killingeq}
so $\Kill^\cI(v)$ is in fact a Killing vector of the field-space manifold at the VEV.

Note that if a tensor is annihilated by $ {\cal L}_\Kill $, then so too is the covariant derivative of that tensor, so in particular
\eq{
 {\cal L}_\Kill {\cal O}_{I_1\cdots I_n}(v) = 0 \qquad \textrm{implies that} \qquad {\cal L}_\Kill \nabla_J {\cal O}_{I_1\cdots I_n}(v) = 0 \, .
}{}
Altogether, this means that the tensor couplings in the Lagrangian as well as their covariant derivatives are all annihilated by $ {\cal L}_\Kill $.  Since on-shell amplitudes are composed precisely out these objects, they are also annihilated, 
\begin{align}
	\label{eq:LieDerAmplitude}
	\mathcal{L}_{\Kill}\Amp{n}^{\fI_{1} \cdots \fI_{n}} =  \Kill_{\fI} \nabla^{\fI} \Amp{n}^{\fI_{1} \cdots \fI_{n}} - \sum^n_{a=1} \nabla_{\fJ_a} \Kill^{\fI_a} \Amp{n}^{\fI_{1} \cdots \fJ_a \cdots \fI_{n}} =0  \, .
\end{align}
This is the geometric statement that scattering exhibits the symmetry.
    
Thus far, we have not drawn any distinctions between linear and non-linear symmetries.  To do so, let us expand 
\Eq{eq:globalsymm} about the VEV using $\Phi^I = v^I + \phi^I$, yielding the symmetry transformation
\eq{
\phi^I \quad \rightarrow \quad \phi^I + \Kill^I(v) +   \partial_J \Kill^I(v)\phi^J + \cdots \, .
}{}
If the symmetry is linear at the VEV, then $\Kill^I(v)=0$ and we identify $\partial_J \Kill^I(v) = T^I{}_J$ as the corresponding generator.   In this case \Eq{eq:LieDerAmplitude} becomes
\eq{
 \sum^n_{a=1} T^{\fI_a}{}_{\fJ_a} \Amp{n}^{\fI_{1} \cdots \fJ_a \cdots \fI_{n}} = 0 \, ,
}{}
which is the usual Ward identity for the amplitude associated with an unbroken symmetry.
    
If, on the other hand, the transformation has an affine component, $\Kill^I(v)\neq 0$, then the symmetry is non-linear.  Moreover, using that the Lie derivative of the covariant derivative of the potential is zero, we find that
\eq{
0= {\cal L}_\Kill V_I(v) = \Kill^J \nabla_J V_I(v) + \nabla_I \Kill^J  V_J(v) = \Kill^J V_{IJ}(v) \,.
}{}
This implies that the Killing vector is a null eigenvector of the mass matrix, and thus defines a massless state.      The massless particle associated with $\Kill^I(v)$ is defined by $\Kill_i(v) |p^i\rangle$.  For obvious reasons, we dub this massless particle a NGB, since this is its identity in theories of spontaneous symmetry breaking.   In the next section, we derive the corresponding soft theorem for this NGB.

\subsection{Nambu-Goldstone Bosons}    
    
We are now ready to study the interplay between symmetry and soft limits.    We restrict to the case of non-linear symmetries in which the spectrum includes a massless NGB.  For simplicity we also assume a vanishing potential.
In this case, the soft limit of the NGB is given by
\begin{align}
	\lim_{q\rightarrow 0} \Kill_{\fI} \Amp{n+1}^{\fI_{1}\cdots \fI_n \fI} = \Kill_{\fI} \nabla^{\fI} \Amp{n}^{\fI_{1}\cdots \fI_{n}} \, ,
\end{align}
which is simply the geometric soft theorem in \Eq{eq:soft_thm} contracted against the Killing vector corresponding to that NGB.

Combining the geometric soft theorem in \Eq{eq:soft_thm} together with the fact that the Lie derivative of the amplitude with respect to the Killing vector is zero in \Eq{eq:LieDerAmplitude}, we obtain\footnote{This formulation of the geometric soft theorem can also be proven directly using current algebra arguments for the Noether current, $ j_\mu{} = \Kill(\Phi)_{\cI}  \partial_\mu \Phi^\cI $, associated with the symmetry transformation.}
\begin{align}
	\lim_{q\rightarrow 0} \Kill_{\fI} \Amp{n+1}^{\fI_{1}\cdots \fI_n \fI}  =  \sum^n_{a=1} \nabla_{\fJ_a} \Kill^{\fI_a} \Amp{n}^{\fI_{1} \cdots \fJ_a \cdots \fI_{n}} \, ,
	\label{eq:softK}
\end{align}
which is a {\it non-zero} soft theorem.  In particular, the soft limit of the $(n+1)$-particle amplitude produces a linear combination of $n$-particle amplitudes.

Non-zero soft theorems have appeared recently in \cite{Kampf:2019mcd} in the context of two-derivative theories of NGBs, and indeed those are a subcase of the soft theorem described above.  Our framework clarifies the content of these non-zero soft theorems by presenting them in terms of geometric quantities which are invariant under changes of field basis. 

A broad class of NGB theories describe the symmetry breaking pattern of a group $G$ to a subgroup $H$. The corresponding effective theory is of course given by a NLSM whose target space is the coset $G/H$. 
Such manifold is endowed with a set of Killing vectors, $\Kill^I_\alpha(\Phi)$ for $\alpha = 1, \cdots, \text{dim}(G)$, which descend from the right-invariant vector fields on the group manifold $G$.  These Killing vectors satisfy
\begin{equation}
[ \Kill_\alpha , \Kill_\beta ]  = f_{\alpha\beta}{}^\gamma \Kill_\gamma\, ,
\end{equation}
which are the commutation relations of the Lie algebra of $G$.

Consider a specific point on $G/H$ labelled by the VEV of the fields, $v^I$.  The full set of Killing vectors $\Kill_\alpha(v)$ can be subdivided into the set of \emph{unbroken} generators $\Kill_a(v)$ and \emph{broken} generators $\Kill_i(v)$.  We define the unbroken generators by
\eq{
\Kill_a(v)=\TKill_a(v) \qquad \textrm{for} \qquad
  a= 1,\cdots,\text{dim}(H) \, ,
 }{}
which satisfy the commutation relations of the Lie algebra of $H$,
\begin{equation}
[ \TKill_a , \TKill_b ]  = f_{ab}{}^c \TKill_c{}  \, .
\end{equation}
Since the unbroken generators stabilize $v^I$, they vanish precisely at that value of the VEV, so
\eq{
\TKill_a^I(v) = 0\,.
}{eq:Tzero}
Note that the identity of the unbroken generators will in general change for different values of the VEV, but the above equation will still hold. Moreover, \Eq{eq:Tzero} only holds at the VEV and not at every point on the manifold, so in general covariant derivatives of the broken generator will not be zero at the VEV.

On the other hand, the broken generators are defined by
\eq{
\Kill_i(v)=\XKill_i(v)  \qquad \textrm{for} \qquad i = 1,\cdots, \text{dim}(G/H) \, ,
}{}
which do not vanish at the VEV and which satisfy the commutation relations
\begin{align}
        [ \TKill_a , \XKill_i ]  &= f_{ai}{}^j \XKill_j  \label{eq:commutatorGH_TX}\,, \\
        [ \XKill_i , \XKill_j ]  &= f_{ij}{}^a \TKill_a + f_{ij}{}^k \XKill_k \,. \label{eq:commutatorGH_XX}
\end{align}
Note that we have used the indices $i,j,k,$ etc.~for the broken generators because they precisely span the tangent space of the field manifold.

At last we are ready to derive a geometric soft theorem for a theory of $G/H$ symmetry breaking.  In particular, we consider \Eq{eq:softK}
where the soft NGB is defined by the Killing vector $\XKill_{\fJ}$ for a broken generator, so
\eq{
        \lim_{q\rightarrow 0} (\XKill_{\fJ})_{\fI} \Amp{n+1}^{\fI_{1} \dots \fI_{n} \fI} = \sum^n_{a=1} \nabla_{\fJ_a} (\XKill_{\fJ})^{\fI_a} \Amp{n}^{\fI_{1} \cdots \fJ_a \cdots \fI_{n}}\,.
}{eq:softX}
In order to simplify the right-hand side we need to compute the quantity $\nabla_{\fJ_a} (\XKill_{\fJ})^{\fI_a}$.
To this end we define $(t_\alpha)_\beta{}^\gamma = f_{\beta\alpha}{}^\gamma$, which is the generator of the adjoint of $G$.
The Killing vectors at an arbitrary point $\Phi^I = v^I + \phi^I$ on the field manifold are related to those at $v^I$ by the adjoint action of $G$ \cite{Boulware:1981ns}, which in components are 
\begin{align}
        \Kill_\alpha^i(\Phi) &= [e^{\phi^j t_i}]_\alpha{}^\beta \, \Kill_\beta{}^i(v)  = [e^{\phi^j t_i}]_\alpha{}^k \XKill_k{}^i(v) \nonumber \\
                            &
                             = \delta_\alpha{}^k \XKill_k{}^i(v) + \phi^i f_{\alpha i}{}^{k} \XKill_k{}^i(v) + \phi^i \phi^j  f_{\alpha i}{}^{\gamma} f_{\gamma j}{}^{k} \XKill_k^i(v) +  \cdots \, ,
\end{align}
where we have used \Eq{eq:Tzero} to drop the unbroken generators, which vanish at the VEV.
We will also need the components of the spin connection, which are given in \cite{Camporesi:1990wm}:
\eq{
        \omega_{ij}{}^k(\Phi) = \tfrac12 f_{ij}{}^k + f_{aj}{}^k \Kill^a{}_i(\Phi)\,.
}{}
Combining the above expressions we find that
\eq{
        \nabla_i (\XKill_j)_k &= \partial_i (\XKill_j)_k + \omega_{ik}{}^l (\XKill_{j})_l  \\
        &= f_{ki}{}^l (\XKill_j)_l + \tfrac12 f_{ik}{}^l(\XKill_j)_l + f_{ak}{}^l \TKill^a{}_i = -\tfrac12 f_{ik}{}^l (\XKill_j)_l  \, ,
}{eq:dkfijk}
where again we have used the vanishing of the broken generators in \Eq{eq:Tzero}.  Plugging \Eq{eq:dkfijk} back into \Eq{eq:softX}, we obtain the geometric soft theorem for a $G/H$ coset space,
\begin{align}
        \lim_{q\rightarrow 0}  \Amp{n+1}^{\fI_{1} \dots \fI_{n} \fI} = -\tfrac12  \sum^n_{a=1} f_{\fJ_a}{}^{\fI_a\fI}  \Amp{n}^{\fI_{1} \cdots \fJ_a \cdots \fI_{n}} \, ,
        \label{eq:softGH}
\end{align}
where we have stripped off the contraction with $\XKill_j$ since it appears on both sides of the soft theorem, which must hold for any $j$. Note that the right-hand side of \Eq{eq:softGH} depends solely on the structure constants of broken generators in \Eq{eq:commutatorGH_XX}.

This result has an interesting geometric interpretation. In addition to the Levi-Civita connection, coset manifolds are endowed with a distinct ``$H$-connection'', $\overline \nabla$, which is also metric compatible but not torsion free \cite{KobayashiNomizu}. The components of the torsion $S= \overline \nabla - \nabla$ are given by the structure constants  \cite{Camporesi:1990wm},
\eq{
S_{i}{}^{jk} = \tfrac12  f_i{}^{kj}\,.
}{}
Thus, when the geometric soft theorem in \Eq{eq:softGH} is non-zero, the right-hand side measures the torsion of the $H$-connection of $G/H$. For a detailed discussion of torsion in NLSM see \cite{Braaten:1985is}.

\subsection{Adler Zero Revisited}

Let us now turn to the scenario in which the right-hand side of the geometric soft theorem is vanishing, {\it i.e.}~exhibits an Adler zero.  We will see how our approach offers a new geometric perspective on the classic soft theorems of the NLSM \cite{Adler:1964um}.

To begin, let us turn to the general two-derivative theory described in Sec.~\ref{subsec:twoder}, assuming an Adler zero condition for all amplitudes.  For the soft limit of the five-particle amplitude in \Eq{GNLSM_amps} this implies that $\nabla^{\fI_5} R^{\fI_1 \fI_2 \fI_3 \fI_4} = 0$, while for six-particle scattering this implies that $\nabla^{(\fI_5}\nabla^{\fI_6)} R^{\fI_1 \fI_2 \fI_3 \fI_4} = 0$.  For $n$-particle scattering, the Adler zero enforces that $\nabla^{(\fI_5} \cdots \nabla^{\fI_n)} R^{\fI_1 \fI_2 \fI_3 \fI_4} = 0$.  In the limit that $n$ goes to infinity, this implies that all symmetric derivatives of the Riemann curvature are zero at the VEV.  This is equivalent to saying that $\nabla^{\fI_5} R^{\fI_1 \fI_2 \fI_3 \fI_4}=0$ at {\it any arbitrary} point in field space, and so the Riemann curvature is covariantly constant. This is possible if and only if the manifold is a locally symmetric space or symmetric coset \cite{Nomizu,Helgason}.  In conclusion, there is an Adler zero \emph{if and only if} the field-space manifold is  symmetric.  The fact that the Adler zero only holds for symmetric cosets is known \cite{Low:2014nga, Cheung:2020tqz}, but our result explains that this is the only class of target space manifolds for which it holds. 

A natural question is then: in what circumstances do higher-derivative deformations of the NLSM preserve the Adler zero \cite{Rodina:2021isd}? Our soft theorem provides a clear answer.  The Adler zero is maintained by any higher-derivative coupling that is a covariantly constant tensor in field space. For instance, the general four-derivative interaction  in the Lagrangian in \Eq{LhigherD} 
only preserves the Adler zero if $\nabla^{\fI_5} \lambda^{\fI_1 \fI_2 \fI_3 \fI_4} = 0$ at all points in field space.

Amusingly, if the field-space manifold is a symmetric space then products of Riemann tensors satisfy additional sets of identities such as
\eq{
	0 = [\nabla^{\fI_1},\nabla^{\fI_2}] R^{\fI_3 \fI_4 \fI_5 \fI_6} = R^{\fI_1 \fI_2 \fI_3}{}_{\fJ}R^{\fJ \fI_4 \fI_5 \fI_6} + R^{\fI_1 \fI_2 \fI_4}{}_{\fJ}R^{\fI_3 \fJ \fI_5 \fI_6} + R^{\fI_1 \fI_2 \fI_5}{}_{\fJ}R^{\fI_3 \fI_4 \fJ \fI_6} + R^{\fI_1 \fI_2 \fI_6}{}_{\fJ}R^{\fI_3 \fI_4 \fI_5 \fJ}\,,
}{eq:Jacobi}
which follow from the fact that the Riemann curvature is covariantly constant.
This reduces the number of independent tensor structures that can be constructed.  For $n$-particle scattering, this number is $(n-2)!$, which exactly coincides with the number of independent color structures in gauge theory \cite{Kleiss:1988ne, DelDuca:1999rs}.  This concordance was required in order for color-kinematics duality and the double copy \cite{Kawai:1985xq, Bern:2008qj,Bern:2010ue,Bern:2019prr} to be applicable to amplitudes in the NLSM on a symmetric coset. 

Yet another way to understand the Adler zero is in terms of Killing vectors. By definition, a symmetric space is a coset whose generators exhibit a $\mathbb{Z}_2$ parity under which the broken generators are odd, so
\eq{
\TKill_a \to \TKill_a \qquad \textrm{and} \qquad \quad \XKill_i \to - \XKill_i\,.
}{}
Compatibility with this parity requires that the commutation relations take the form
\begin{align}
        [ \TKill_a , \XKill_i ]  &= f_{ai}{}^j \XKill_j  \label{eq:commutatorGHsym_TX}\,, \\
        [ \XKill_i , \XKill_j ]  &= f_{ij}{}^a \TKill_a \, , \label{eq:commutatorGHsym_XX}
\end{align}
which is to say that  $f_{ij}{}^k=0$ in \Eq{eq:commutatorGH_XX}.
As a consequence, the covariant derivatives of  broken generators are zero \cite{Eschenburg},  
\eq{
\nabla_\fI \Kill_\fJ{}^I (v)  = \nabla_\fI \XKill_\fJ{}^I (v) = 0\,,
}{eq:dkvzero} 
which can be shown using \Eq{eq:dkfijk}. Using \Eq{eq:softK} or \Eq{eq:softGH}, we see that right-hand side of the soft theorem vanishes for symmetric spaces, thus yielding the Adler zero. 

Our reformulation of the Adler zero in terms of Killing vectors reveals an unnecessary assumption in the usual current algebra proof of the Adler zero for two-derivative theories without a potential \cite{Adler:1964um}.  In particular, it is {\it not necessary} to require the absence of three-point interactions and linear terms in the symmetry transformation. So their presence cannot ever preclude a vanishing soft limit.  Indeed, by solving the Killing equation \eqref{eq:killingeq} as a power series about the VEV and using \Eq{eq:dkvzero},  
\eq{
     \Kill_\cI(\Phi) &= \Kill_\cI(v) +  \Gamma^\cJ{}_{\cI\cK}(v) \Kill_\cJ(v) \phi^\cK + \cdots\, ,
}{}
and also noting that
\eq{
g_{\cI\cJ}(\Phi)  &= g_{\cI\cJ}(v)  + 2\,\Gamma_{(\cI\cJ)\cK}(v)  \, \phi^K + \cdots \, ,
}{}
we see that the linear term of the symmetry transformation \Eq{eq:globalsymm} and the three-point two-derivative interaction are both related to the Christoffel symbol. In fact, the relation is such that their effects precisely cancel in an amplitude, which can be easily confirmed by performing a field redefinition to Riemann normal coordinates, where the Christoffel symbol is zero. This simultaneously eliminates both the cubic interaction and the linear term in the symmetry transformation.
In hindsight this had to be true given that the presence of three-point interactions is a field-basis-dependent statement, whereas the Adler zero is not.

Last but not least, it is natural to ask whether there are alternative ways that an Adler zero can arise on the right-hand side of \Eq{eq:softK}.   For example, is it possible to have $        \nabla_{\fJ} \Kill^{\fI} = 0$ at {\it all} points in field space? This is equivalent to asking whether the Killing vector associated with the NGB can be covariantly constant.  While this is possible, it is simple to show that such a degree of freedom couples rather trivially.  Since the Lie derivative ${\cal L}_\Kill$ annihilates all coupling tensors and their covariant derivatives, then so too does $n$ powers of the Lie derivative, which for covariantly constant $\Kill$ is equal to ${\cal L}_\Kill^n= \Kill^{i_1}\cdots \Kill^{i_n} \nabla_{(i_1} \cdots  \nabla_{i_n)}$.  The fact that every coupling is annihilated by ${\cal L}_\Kill^n$ implies that there is a field basis in which the coupling tensors have no dependence on the corresponding NGB.  So for example in the Lagrangian in \Eq{L_general}, the couplings $g_{IJ}$, $V$, and $\lambda_{IJKL}$ would have no dependence on the NGB, although the NGB could still appear derivatively coupled through $\partial_\mu \Phi^I$.

\subsection{Dilatons}

The soft dilaton theorem \cite{Callan:1970yg,Boels:2015pta,Huang:2015sla ,DiVecchia:2015jaq} is actually
a corollary of our geometric soft theorem.   To understand why, let us construct the effective Lagrangian for a dilaton coupled to an arbitrary theory of scalar ``matter'' fields.  By definition, the dilaton is a compensator for scale transformations, so one can construct its interactions via the so-called Stueckelberg trick.  In particular, starting from the scalar matter field Lagrangian, we apply a scale transformation and then promote the scale parameter to a dynamical dilaton.

To start, consider a theory of scalar matter fields described by \Eq{L_general}.  Previously, we expanded $\Phi^I = v^I + \phi^I$ about an arbitrary VEV.  In the case of the dilaton this is not permitted, however.  The reason for this is that the dilaton is by definition the compensator for {\it all} scales, including VEVs.  Said another way, if there are any fields in the theory that acquire VEVs other than the dilaton, then these states will necessarily mix with the dilaton.  To simplify our analysis, we eliminate this mixing from the start by assuming $v^I = 0$ for the scalar matter fields.

In this basis the scalar matter field Lagrangian is
\eq{
L = \tfrac{1}{2} g_{IJ}(\phi) \partial_\mu \phi^I \partial^\mu \phi^J -V(\phi) + \cdots \, ,
}{eq:L_matt}
where the ellipses denote higher-derivative interactions.  Next, consider a scale transformation,
\eq{
x^\mu \quad \to \quad  \xi^{{1}/{\Delta}}  x^\mu \qquad \textrm{and} \qquad \phi^{ I} \quad \to \quad \dil^{-1} \phi^{ I}  \, ,
}{eq:scale_transform}
where by convention we have chosen $\xi$ to have the scaling dimension $\Delta = \frac{D-2}{2}$ of a scalar in $D$ dimensions. Applying \Eq{eq:scale_transform} 
to \Eq{eq:L_matt} and promoting the scale parameter $\dil$ to a dynamical dilaton, 
we obtain the dilaton effective Lagrangian,
\eq{
L &= \tfrac{1}{2} \partial_\mu \dil \partial^\mu \dil + \tfrac{1}{2} g_{IJ}(\dil^{-1} \phi) \partial_\mu \phi^I \partial^\mu \phi^J - \dil^{D/\Delta}V(\dil^{-1}\phi) + \cdots \, .
}{}
Note that the dilaton kinetic term is fixed by its scaling dimension.   Naively, the appearance of inverse powers of the dilaton may seem peculiar but this is not an issue because the VEV of the dilaton, $\langle \dil\rangle$, is assumed to be non-zero.

Let us now study the dilaton effective Lagrangian expanded about the vanishing VEVs of the scalar matter fields.  First, we consider the two-derivative terms,
 \eq{
\tfrac{1}{2} g_{IJ}(\dil^{-1} \phi) =  \tfrac{1}{2} (g_{IJ}(0)+\partial_K g_{IJ}(0)\dil^{-1} \phi^K+ \cdots) \partial_\mu \phi^I \partial^\mu \phi^J  \, .
}{}
Crucially, the lowest order coupling of the dilaton is to {\it three} scalar matter fields.  Hence, the theory includes a dilaton-matter-matter-matter coupling but no dilaton-matter-matter coupling.  The absence of the latter then implies that any Christoffel symbol involving the dilaton---which encodes the corresponding cubic vertices---is zero.  This implies the geometric statement,
\eq{
\nabla_{\langle \dil \rangle} = \partial_{\langle \dil \rangle} \, ,
}{}
so covariant and partial derivatives with respect to the VEV of the dilaton are one and the same. More invariantly, this means that the dilaton parameterizes a flat direction in field space that non-linearly realizes scale transformations.

Second, let us consider the expansion of the potential terms about the VEVs, yielding
\eq{
\dil^{D/\Delta}V(\dil^{-1}\phi) =  \cdots + \tfrac{1}{2} \dil^{2/\Delta} \partial_I \partial_J V(0)  \phi^I \phi^J  +\cdots \, .
}{}
Here we see that the mass term for the scalar matter field is simply dressed by a factor of $\dil^{2/\Delta}$.   This is of course expected, since the dilaton modulates the mass spectrum uniformly.  It will be useful later to realize then that
\eq{
\nabla_{\langle \dil \rangle } V_{IJ} =  \tfrac{2}{\Delta \langle \xi\rangle } V_{IJ} \, ,
}{}
so the covariant derivative of the mass matrix with respect to the dilaton VEV acts as a simple multiplicative factor.

At last, we are now equipped to apply the geometric soft theorem in \Eq{eq:soft_thm} to the case of the dilaton.  The soft limit of an amplitude with a single dilaton and $n$ scalar matter fields is
\eq{
  \lim_{q\rightarrow 0} \Amp{n+1}^{i_1\cdots i_n  \langle \dil \rangle} 
  &= \nabla^{\langle \dil \rangle} \Amp{n}^{i_1\cdots i_n} 
 +  \sum_{a=1}^n 
  \frac{ \nabla^{\langle \dil \rangle} V_{j_a}^{i_a  }}{(p_a +q)^2  - m_{j_a}^2} \left(1+ q^\mu \frac{\partial}{\partial p_a^\mu} \right)
\Amp{n}^{i_1 \cdots j_a \cdots i_n} \\
& =  \partial_{\langle \dil \rangle} \Amp{n}^{i_1\cdots i_n} 
 + \tfrac{1}{\Delta \langle \xi \rangle} \sum_{a=1}^n 
  \frac{m_{i_a}^2}{p_a\cdot q} \left(1+ q^\mu \frac{\partial}{\partial p_a^\mu} \right)
\Amp{n}^{i_1 \cdots i_a \cdots i_n} \, .
}{eq:soft_dilaton_thm}
The above expression is secretly identical to the dilaton soft theorem described in  \cite{Callan:1970yg, Boels:2015pta, Huang:2015sla, DiVecchia:2015jaq}.  To understand why, recall that dimensional analysis implies that
\eq{
\left(\Delta \langle\dil\rangle \partial_{\langle\dil\rangle} + \sum_{a=1}^n p_a^\mu \frac{\partial}{\partial p_a^\mu}   \right) \Amp{n}^{i_1 \cdots i_n} = (D- n\Delta)\Amp{n}^{i_1 \cdots  i_n}\, .
}{eq:dim_analysis}
The left-hand side extracts the total mass dimension of the amplitude by counting the overall powers of momenta and dimensionful coupling constants.  By definition, the latter enter everywhere with factors of the dilaton VEV.    Plugging \Eq{eq:dim_analysis} into \Eq{eq:soft_dilaton_thm} to eliminate $\partial_{\langle \dil \rangle}$, we obtain the soft dilaton theorem in its more standard form,
\eq{
  \lim_{q\rightarrow 0} \Amp{n+1}^{i_1\cdots i_n  \langle \dil \rangle} 
& = \frac{1}{f} \left[D- n\Delta - \sum_{a=1}^n \left( p_a^\mu \frac{\partial}{\partial p_a^\mu} +
  \frac{m_{i_a}^2}{p_a\cdot q} \left(1+ q^\mu \frac{\partial}{\partial p_a^\mu} \right) \right) \right]
\Amp{n}^{i_1 \cdots i_a \cdots i_n} \, ,
}{}
where we have defined the decay constant of the dilaton, $f = \Delta \langle \dil \rangle$ \cite{DiVecchia:2017uqn}.

\section{Multiple-Soft Theorems}

\label{sec:multisoft}

In this section we consider multiple-soft theorems that govern the behavior of amplitudes as a set of particles are taken soft, either consecutively or simultaneously.  In the former, the soft particles exhibit a hierarchy of softness that dictates their ordering.  In the latter, all soft particles are on the same footing.  Throughout this analysis we focus on the leading non-trivial order, dropping contributions ${\cal O}(q)$ and higher.
Furthermore, for simplicity we consider scalar field theories in which the potential is vanishing.

\subsection{Consecutive Double-Soft Theorem}

The consecutive double-soft theorem is trivially obtained by applying the geometric soft theorem in \Eq{eq:soft_thm} on a pair of particles in sequence.  A quantity of particular interest is the {\it commutator} of consecutive double-soft limits, which a trivial calculation shows to be
\eq{
    \left[\lim_{q_a\to 0}, \lim_{q_b\to 0}\right]\Amp{n+2}^{i_1 \cdots i_n i_a i_b}
&= [\nabla^{\fI_a},\nabla^{\fI_b}]\Amp{n}^{i_1 \cdots i_n}  = \sum _{c\neq a,b}R^{\fI_{a}\fI_{b}\fI_{c}}{}_{\fJ_c} \Amp{n}^{i_1 \cdots j_c \cdots  i_n} \,  .
}{eq:consDoubleSoft}
The above expression is beautifully intuitive, since the commutator of soft limits measures the change of the amplitude when transported around an infinitesimal square in field space---thus probing the local curvature.  Note also that \Eq{eq:consDoubleSoft} holds generally, {\it i.e.}~for arbitrary two- and higher-derivative interactions.   

For the special case of the NLSM on a symmetric manifold, the right-hand side of \Eq{eq:consDoubleSoft} sums to zero by the Jacobi identities, which are automatically satisfied by the Riemann curvature on account of \Eq{eq:Jacobi}.  This is necessary for consistency with the Adler zero in the single-soft limit, which directly implies that the consecutive double-soft limit is also vanishing.

\subsection{Simultaneous Double-Soft Theorem}

Next, let us consider the simultaneous double-soft theorem, whereby a pair of particles is taken soft at the same rate.  Such a setup has previously been studied in the NLSM \cite{Weinberg:1966kf,Arkani-Hamed:2008owk,Cachazo:2015ksa,Du:2015esa,Low:2015ogb}, both at leading and subleading order. Here we prove that the leading double-soft limit for a general scalar theory is
\eq{
     \lim_{q_a, q_b\to 0}\Amp{n+2}^{i_1 \cdots i_n i_a i_b}
	&= \tfrac12 \sum _{c\neq a,b}   \frac{s_{ac}-s_{bc}}{s_{ac}+s_{bc}} R^{\fI_{a}\fI_{b}\fI_{c}}{}_{\fJ_c} \Amp{n}^{i_1 \cdots j_c \cdots  i_n} +  \nabla^{(\fI_a}\nabla^{\fI_b)}  \Amp{n}^{i_1 \cdots i_n}    \,.
}{eq:doubleSimSoft}
The first term already appears in the NLSM but the second term appears for a general theory without a potential.  One can verify directly that this double-soft theorem holds for the amplitudes in Secs.~\ref{subsec:twoder} and \ref{subsec:lambda}.

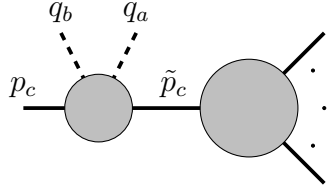
\begin{figure}
    \centering
\begin{tikzpicture}
            \coordinate (c) at (-3, 0);	
	    	\coordinate (b) at (-2.5, 1);
	    	\coordinate (a) at (-1.5, 1);
	    	\node[above]  at (a) {$q_a$};
	        \node[above]  at (b) {$q_b$};
	        \node[above]  at (c) {$p_c$};
	        \node[above]  at (-1, 0 ) {$\tilde p_c$};
	    	
	    	\coordinate (f) at (1, 1);
	    	\coordinate (l) at (1, -1);
	    	
	    	\coordinate (m4) at (-2,0);
	    	\coordinate (mn) at (0, 0);
	    		    	\node at (mn) {$ \Amp{n}$};
	    	\node at (m4) {$\tilde A_4$};
	    	
	    	\draw [hard] (c) -- (m4);
	    	\draw [hard] (mn) -- (m4);
	    	\draw [soft] (b) -- (m4);
	    	\draw [soft] (a) -- (m4);
	    	
	    	\draw [hard] (f) -- (mn);
	        \draw [hard] (l) -- (mn);
	    
	    	\coordinate (d1) at (1.00, 0);
	    	\coordinate (d2) at (0.85, 0.5);
	    	\coordinate (d3) at (0.85, -0.5);

            \draw[fill=lightgray, opacity=1] (mn) circle (0.65);
            \draw[fill=lightgray, opacity=1] (m4) circle (0.45);

	\draw[fill=black, opacity=1] (d1) circle (0.02);
            \draw[fill=black, opacity=1] (d2) circle (0.02);
            \draw[fill=black, opacity=1] (d3) circle (0.02);
\end{tikzpicture}
    \caption{Diagram containing potentially singular terms in the simultaneous double-soft theorem.}
    \label{fig:sing_doublesoft}
\end{figure}

Our proof of \Eq{eq:doubleSimSoft} is a generalization of the analysis in \cite{Low:2015ogb} for the NLSM. The amplitude is organized into potentially singular terms, which have diagrammatic structure as in Fig.~\ref{fig:sing_doublesoft}, and regular terms, which we denote by $\cal R$, as follows
\begin{align}
        \Amp{n+2}^{\fI_{1} \cdots \fI_{n} i_a i_b}(p_1, \cdots, p_n, q_a, q_b) =& \phantom{{}+{}} \sum_{c\neq a,b} \Amp{4}^{i_a i_b i_c }{}_{j_c}(q_a, q_b,p_c,  -\tilde p_c) \frac{1}{\tilde{p}_c^2} \Amp{n}^{\fI_{1} \cdots j_c \cdots \fI_{n} }( p_1, \cdots , \tilde p_c, \cdots p_n)  \nonumber \\
      &  + {\cal R}^{\fI_{1} \cdots \fI_{n} i_a i_b}(p_1, \cdots, p_n, q_a, q_b)\,, \label{eq:dssplit}
\end{align}
where $\Amp{n}$ is the amputated $n$-particle current with the momentum $\tilde p_c = p_c -q_a -q_b$ taken off-shell.  The function ${\cal R}$ is a remainder function that is local in the soft momenta.
By explicit calculation we find that the four-particle current is
\eq{
\Amp{4}^{i_a i_b i_c }{}_{j_c} =&
 \phantom{{}+{}} 
 \tfrac12 R^{\fI_a\fI_b\fI_c }{}_{\fJ_c} ( s_{ac} - s_{bc})
+
 \tfrac16 (R^{ \fI_c \fI_a  \fI_b }{}_{\fJ_c} + R^{\fI_c \fI_b\fI_a}{}_{\fJ_c}) ( s_{ac} + s_{bc} - 2s_{ab}) \\
 &  +  \tfrac{1}{2} s_{ab} s_{c\tilde c} \lambda^{\fI_a \fI_b \fI_c}{}_{\fJ_c} + \tfrac{1}{2} s_{ac} s_{b\tilde c} \lambda^{\fI_a \fI_c \fI_b}{}_{\fJ_c} + \tfrac{1}{2} s_{bc} s_{a\tilde c} \lambda^{\fI_b \fI_c \fI_a}{}_{\fJ_c}
         \,,
}{eq:A4}
as shown in Fig.~\ref{fig:sing_doublesoft}. Here we have explicitly included the four-derivative coupling, $\lambda$, but all of our results will equally apply when including other higher-derivative interactions.
Dropping contributions subleading in the soft expansion, we can effectively set $s_{ab}=s_{a\tilde c}=s_{b\tilde c}=s_{c\tilde c}=0$ and $\tilde p_c^2= s_{ac}+s_{bc}$, while setting the momenta $\tilde p_c$ in the subamplitudes equal to $p_c$. Combining \Eq{eq:A4} with \Eq{eq:dssplit} we obtain 
\begin{align}
    \lim_{q_a,q_b\to 0}     \Amp{n+2}^{\fI_{1} \cdots \fI_{n} i_a i_b} =& \phantom{{}+{}} \tfrac{1}{2} \sum_{c\neq a,b}  \frac{s_{ac}-s_{bc}}{s_{ac}+s_{bc}} R^{\fI_a \fI_b \fI_c}{}_{\fJ_c}  \Amp{n}^{\fI_{1} \cdots j_c \cdots \fI_{n} } \nonumber \\
     &+\tfrac{1}{6} \sum_{c\neq a,b}  (R^{ \fI_c  \fI_a \fI_b}{}_{\fJ_c} + R^{\fI_c \fI_b\fI_a}{}_{\fJ_c})   \Amp{n}^{\fI_{1} \cdots j_c \cdots \fI_{n} } 
        +   \lim_{q_a,q_b\to 0}    {\cal R}^{\fI_{1} \cdots \fI_{n} i_a i_b} \, . \label{eq:dsstep}
\end{align}
Note that the four-derivative coupling, $\lambda$, does not affect the result at this order. The same is true of other higher-derivative couplings.
To simplify the above expression, we compute the {\it anti-commutator} of consecutive soft limits of \Eq{eq:dssplit}, yielding
\eq{
\tfrac12 \left\{ \lim_{q_a\to 0}, \lim_{q_b\to 0} \right\} \Amp{n+2}^{\fI_{1} \cdots \fI_{n} i_a i_b} &=  \tfrac16 \sum_{c\neq a,b} (R^{  \fI_c \fI_a \fI_b}{}_{\fJ_c} + R^{\fI_c \fI_b\fI_a}{}_{\fJ_c})   \Amp{n}^{\fI_{1} \cdots j_c \cdots \fI_{n} }+  \lim_{q_a,q_b\to 0}    {\cal R}^{\fI_{1} \cdots \fI_{n} i_a i_b} \,.
}{eq:anticomm1}
We now compare this to the anti-commutator of consecutive soft limits computed directly from our geometric soft theorem, 
\eq{
\tfrac12 \left\{ \lim_{q_a\to 0}, \lim_{q_b\to 0} \right\} \Amp{n+2}^{\fI_{1} \cdots \fI_{n} i_a i_b} &= \nabla^{(\fI_a}\nabla^{\fI_b)}A_{n}^{i_1 \cdots i_n}\,.
}{eq:anticomm2}
Combining \Eq{eq:dsstep} with \Eq{eq:anticomm1} and \Eq{eq:anticomm2}, we obtain the claimed simultaneous double-soft theorem in \Eq{eq:doubleSimSoft}.

\subsection{Simultaneous Triple-Soft Theorem}

It is straightforward to generalize our proof of the simultaneous double-soft theorem to a triplet of particles.
Applying parallel reasoning, we can prove a new simultaneous triple-soft theorem,
\eq{	    
 \lim_{q_a, q_b,q_c\to 0} & \Amp{n+3}^{i_1 \cdots i_n i_a i_b i_c}
	 \\ 
	=& \sum _{d\neq a,b,c}\left( \frac{1}{2}  \frac{s_{ad} - s_{bd}}{s_{ad} + s_{bd}} R^{\fI_a \fI_b \fI_d}{}_{\fJ_d}  \nabla^{\fI_c} + \frac{1}{3}\frac{s_{ad} - s_{bd}}{s_{ad} + s_{bd}+s_{cd}} \nabla^{\fI_c}R^{\fI_a \fI_b \fI_d}{}_{\fJ_d}\right)
	\Amp{n}^{\fI_{1} \cdots j_d \cdots \fI_{n} }
         \\ 
        &+
	\frac{1}{3} \frac{s_{ac} - s_{bc}}{s_{ab} + s_{ac}+ s_{bc}} R^{\fI_a \fI_b \fI_c}{}_{\fJ_d}  \nabla^{\fJ_d}
	\Amp{n}^{\fI_{1} \cdots \fI_{n} }
	+ (a \leftrightarrow c) + (b \leftrightarrow c)
	+
	\nabla^{(\fI_a}\nabla^{\fI_b}\nabla^{\fI_c)} \Amp{n}^{\fI_{1} \cdots j_d \cdots \fI_{n} }
	 \, ,
}{}
which is applicable to any theory without a potential.  Note that this formula has no analog in the NLSM, for which all amplitudes with odd numbers of particles are trivially zero.  We have checked explicitly that this triple-soft theorem holds for seven-particle amplitudes in theories without a potential.

\section{On-Shell Recursion Relations}

\label{sec:recursion}

It has long been appreciated that the Adler zero condition can be leveraged to bootstrap the scattering amplitudes of the NLSM \cite{Susskind:1970gf,Osborn:1969ku}.  More recently, similar logic has been applied to a broader class of scalar effective field theories which exhibit enhanced symmetries \cite{Cheung:2014dqa, Cheung:2016drk}, including Dirac-Born-Infeld (DBI) theory and the special Galileon (SG) theory \cite{Hinterbichler:2015pqa}.  Notably, this soft bootstrap can be systematized using on-shell recursion relations \cite{Cheung:2015ota} that directly exploit the soft behavior of amplitudes.   See \cite{Kampf:2013vha, Luo:2015tat,Elvang:2017mdq,Elvang:2018dco,Cheung:2018oki,Low:2019ynd,Kampf:2021bet,Kampf:2021tbk} for other recent work on the soft bootstrap.

These earlier explorations all suggest an intimate correlation between enhanced symmetry, vanishing soft limits, and on-shell constructibility.  In the present work, we have shown that completely generic theories exhibit a geometric soft theorem even in the absence of symmetry.  Interestingly, in many cases the soft limit does not even vanish.  As we will see, we can still exploit the geometric soft theorem to build an on-shell recursion relation for any arbitrary two-derivative scalar field theory.  A priori, it is rather unintuitive that such generic theories should be on-shell constructible.  However, one should bear in mind that the extra amplitudes data that is implicitly input into the recursion is the behavior of amplitudes in a small neighborhood about the VEV.

Before we begin, let us review the original Britto-Cachazo-Feng-Witten (BCFW) recursion relations \cite{Britto:2004ap,Britto:2005fq}.  We deform a pair of external momenta,
\eq{
	p_{1} \rightarrow p_{1} + z q \qquad {\rm and} \qquad p_{2} \rightarrow p_{2} - z q  \, ,
}{}
where $z$ is a complex number and $q$ is reference momentum satisfying $q \cdot p_1  = q\cdot p_2= q^2 =0$. We then apply Cauchy's theorem to write the on-shell $n$-particle scattering amplitude as
\eq{
	\Amp{n}(0) = \frac{1}{2\pi i} \oint \frac{dz}{z} \Amp{n}(z) = -\sum_{\alpha} {\rm Res}_{z=z_{\alpha}} \left(\frac{\Amp{n}(z)}{z} \right)  \, .
}{}
The right-hand side is composed of a sum of residues corresponding to on-shell factorization diagrams and a contribution at $z=\infty$. If the residue at $z=\infty$ is zero then the $A_{n}(0)$ can be calculated recursively from on-shell lower-particle amplitudes.

Theories with derivative interactions typically have poor high energy behavior, in which case the residue at $z=\infty$ will generically non-zero. Nevertheless, the residue at infinity can actually be {\it eliminated} if the soft limit is known \cite{Cheung:2015ota}. To do so, we define a soft momentum shift,
\eq{
	p_{a} \rightarrow p_{a} \left( 1-z c_{a} \right)  ,
}{soft_shift}
with the condition that the deformed momenta should maintain total momentum conservation.\footnote{For $n>D+1$ it is possible to find distinct $c_{i}$ for generic $p_{i}$.} Now we apply Cauchy's theorem to a slightly modified integrand, 
\eq{
	\Amp{n}(0) = \frac{1}{2\pi i} \oint \frac{dz}{z} \frac{\Amp{n}(z)}{F_{n,m}(z)} = - \sum_{\alpha} {\rm Res}_{z=z_{\alpha}^\pm} \left( \frac{\Amp{n}(z)}{z F_{n,m}(z) } \right) ,
}{eq:softrec}
where we have defined the form factor
\eq{
	F_{n,m} (z) = \prod^{n}_{a=1} (1 - c_{a} z)^{m} \, .
}{}
In the case of a vanishing soft limit, the the poles generated by $F_{n,m}(z)$ in the denominator are cancelled by the zeros in the amplitude.  Hence, the only residues which contribute arise from factorization channels, each with a residue equal to the product of lower-particle amplitudes as dictated by unitarity
\eq{
\Amp{n}(z) = \frac{A_{L}(z) A_R(z)}{P_\alpha^2(z)} + \cdots \, ,
}{}
where the propagator denominator is
\eq{
P_\alpha^2(z) = P_\alpha^2 + 2 z P_\alpha\cdot Q_\alpha + z^2 Q_\alpha^2  = P_\alpha^2\,  (1-z/z^+_\alpha)(1-z/z^-_\alpha)\,,
}{}
where $P_\alpha = \sum_a p_a$ and $Q_\alpha = - \sum_a c_a p_a $.  Here $z_\alpha^\pm$ are the two roots of the quadratic polynomial $P_\alpha^2(z)$. 
Thus the resulting recursion formula is
\eq{
	\Amp{n}(0) =  \sum_{\alpha}   \frac{A_{L}(z_\alpha^+) A_R(z_\alpha^+)}{(1-z_\alpha^+/z_\alpha^-) F_{n,m}(z_\alpha^+) } + (z_\alpha^+ \leftrightarrow z_\alpha^-) \, ,
}{}
and can be used to construct all scattering amplitudes in the NLSM, DBI, and SG.

On the other hand, when the soft limit of the amplitude is known but non-zero, then one can construct a generalized version of soft on-shell recursion relations \cite{Luo:2015tat}. We instead consider 
\begin{align}
        \Amp{n}(0) &= \frac{1}{2\pi i} \oint \frac{dz}{z} \frac{\Amp{n}(z)}{F_{n,1}(z)} \nonumber \\
        &= - \sum_{\alpha} {\rm Res}_{z=z_{\alpha}^\pm} \left( \frac{\Amp{n}(z)}{z F_{n,1}(z) } \right) - \sum_{a} {\rm Res}_{z=1/c_a} \left( \frac{\Amp{n}(z)}{z F_{n,1}(z) } \right)  ,
	\label{eq:softrecsub}
\end{align}
where in addition to the factorization poles we need to include the additional residues at $z=1/c_a$ from the poles in $F_{n,1}(z)$.

At last, we are equipped to use our geometric soft theorem to derive on-shell recursion relations.  For simplicity, let us consider a general two-derivative theory of scalars without a potential, as defined in \Eq{L_GNLSM}.  Crucially, in such a theory the soft limit need not vanish.   In this case, each residue in the last term of \Eq{eq:softrecsub} becomes
\eq{
        {\rm Res}_{z=1/c_a} \left( \frac{\Amp{n}(z)}{z F_{n,1}(z) } \right)= \frac{\nabla_{i_a} A_{n-1}(1/c_a)}{\Pi_{b\neq a} ( 1 - c_{b}/c_{a})}\,.
}{}
This yields the on-shell recursion relation
\eq{
	\Amp{n}(0) =  \sum_{\alpha}   \frac{A_{L}(z_\alpha^+) A_R(z_\alpha^+)}{(1-z_\alpha^+/z_\alpha^-) F_{n,1}(z_\alpha^+) } + (z_\alpha^+ \leftrightarrow z_\alpha^-) + \sum_a \frac{\nabla_{i_a} \Amp{n-1}(1/c_a)}{\Pi_{b\neq a} ( 1 - c_{b}/c_{a})} \, ,
}{eq:recursion_two_deriv}
which is valid for a general two-derivative scalar theory.

Now let us check \Eq{eq:recursion_two_deriv} for the very simplest case of building the five-particle amplitude in terms of the four-particle amplitude via on-shell recursion.  
To more easily construct the momentum shift in \Eq{soft_shift} we specialize to three-dimensional kinematics.  Note that this choice actually introduces no loss of generality, since the amplitudes of two-derivative scalar theories are degree one polynomials in Mandlestam variables, so they cannot include a three-dimensional Gram determinant.   An explicit solution for the $c_{a}$ is \cite{Luo:2015tat}
\begin{align}
	c_{1} &= s_{23} (s_{14} s_{23} - s_{13} s_{24} - s_{12} s_{34}) \,, \nonumber \\
	c_{2} &= s_{13} (- s_{14} s_{23} + s_{13} s_{24} - s_{12} s_{34})\,, \nonumber \\
	c_{3} &= s_{12} (- s_{14} s_{23} - s_{13} s_{24} + s_{12} s_{34})\,, \\
	c_{4} &= 2 s_{12} s_{13} s_{23}\,, \nonumber \\
	c_{5} &= 0 \, . \nonumber 
\end{align}
There are no contributions to \Eq{eq:recursion_two_deriv} from factorization channels because the three-particle amplitude vanishes.  Thus, the  five-particle amplitude arises purely from the residues $z\rightarrow 1/c_{a}$. Explicitly, the five-particle amplitude is
\eq{
	\Amp{5}(0) &= \sum_{a=1}^{4} \frac{ \nabla_{i_a} \Amp{4,a}(1/c_{a})}{\Pi^{4}_{b=1,b\neq a} ( 1 - c_{b}/c_{a})} \, ,
}{eq:A5_recursion}
where $\Amp{4,a}(1/c_{a})$ is the four-particle amplitude not including particle $a$, and with momenta shifted by $z\rightarrow 1/c_{a}$. It is straightforward to check that \Eq{eq:A5_recursion} agrees precisely with the five-particle scattering amplitude we computed previously in \Eq{GNLSM_amps}. We have also checked numerically that  \Eq{eq:recursion_two_deriv} produces the correct six-particle amplitude.

Via repeated application of these on-shell recursion relations, any $n$-particle amplitude can be expressed in terms of the four-particle amplitude in \Eq{GNLSM_amps}.  Again, it may seem peculiar that this is even possible, since a general two-derivative scalar theory does not exhibit any enhanced symmetry.  However, the key point here is that the four-particle amplitude depends on the Riemann curvature as a function of the VEV.  This quantity, together its covariant derivatives, encode all of the properties of the underlying manifold.  So implicitly, the seed of the on-shell recursion relation is not simply the four-particle amplitude, but the four-particle amplitude evaluated at {\it all points} on the manifold.

\section{Conclusions}

In this paper we have described a systematic framework for recasting the dynamics of scattering in terms of the geometry of field space.  On-shell scattering amplitudes are physical quantities, so they must be invariant under changes of field basis.  Hence, they can be expressed in terms of a set of corresponding geometric invariants.   In particular, we have shown quite generally how a variety of geometric objects are actually avatars for familiar physical concepts, including LSZ reduction (the tetrad), Ward identities (the Lie derivative), single-soft limits (the covariant derivative), and multiple-soft limits (the curvature).

Regarding soft structure, we have shown how the geometry of field space mandates a universal soft theorem applicable to any theory of scalars, even including masses, potential interactions, and higher-derivative couplings.  In general, the soft limit is non-vanishing.  For theories with symmetry, we have shown how the soft theorem is exactly dictated by the associated Killing vectors in field space. When the symmetries are affine, the corresponding theories describe NGBs, and we show under what circumstances there exists an analog of the Adler zero.  We also show how the dilaton soft theorem is another corollary of our geometric soft theorem.  Last but not least, we have also derived new double- and triple-soft theorems and showed how to leverage our results to derive on-shell recursion relations that implement the soft bootstrap for a much broader class of theories.

The present work offers numerous avenues for future analysis.  First and foremost is the generalization of our results to an even broader class of theories.  In particular, it is obvious that the geometry of scalar field space will persist even with inclusion of spectator particles with spin, such as fermions and vectors.  Hence, the soft limits of scalars will be universal in those theories as well.  On the other hand, a more ambitious goal would be to apply our reasoning to the soft limit of particles with spin, as in the Weinberg soft theorems for gauge theory and gravity, and in non-relativistic or condensed matter systems. 

Second, it would be interesting to understand the broader class of invariances under {\it derivative} field redefinitions.  In particular, it is known that on-shell scattering amplitudes are unchanged by changes of field basis involving derivatives.  What are the geometric invariants associated with these changes of basis? To our knowledge, there exists no unified framework to understand this structure.

Third, there is the question of further applications for scattering amplitudes.  A natural question is whether there exist subleading generalizations of our geometric single- and multiple-soft theorems.    Another question is whether one can construct on-shell recursion relations that leverage both kinematics as well as the geometry of field space.  In such a construction, one would shift not only the external kinematics of an amplitude, but {\it also} the VEVs that dictate the couplings and masses that appear within it.

\newpage 
\begin{center} 
   {\bf Acknowledgments}
\end{center}
\noindent 
We are grateful to Lance Dixon, Maria Derda, Aneesh Manohar, and Ira Rothstein for useful discussions and comments on the paper. 
We also thank Maria Derda, Elizabeth Jenkins, Aneesh Manohar, and Michael Trott for collaboration on related projects.
C.C., A.H.,~and J.P.-M. are supported by the DOE under grant no.~DE-SC0011632 and by the Walter Burke Institute for Theoretical Physics.

\bibliographystyle{utphys-modified.bst}
\bibliography{bibliographySoftTheorem}

\providecommand{\href}[2]{#2}\begingroup\raggedright\begin{thebibliography}{10}

\bibitem{Low:1958sn}
F.~E. Low, ``{Bremsstrahlung of very low-energy quanta in elementary particle
  collisions},'' \href{http://dx.doi.org/10.1103/PhysRev.110.974}{{\em Phys.
  Rev.} {\bfseries 110} (1958) 974--977}.

\bibitem{Weinberg:1965nx}
S.~Weinberg, ``{Infrared photons and gravitons},''
  \href{http://dx.doi.org/10.1103/PhysRev.140.B516}{{\em Phys. Rev.} {\bfseries
  140} (1965) B516--B524}.

\bibitem{Burnett:1967km}
T.~H. Burnett and N.~M. Kroll, ``{Extension of the low soft photon theorem},''
  \href{http://dx.doi.org/10.1103/PhysRevLett.20.86}{{\em Phys. Rev. Lett.}
  {\bfseries 20} (1968) 86}.

\bibitem{Bern:2014vva}
Z.~Bern, S.~Davies, P.~Di~Vecchia, and J.~Nohle, ``{Low-Energy Behavior of
  Gluons and Gravitons from Gauge Invariance},''
  \href{http://dx.doi.org/10.1103/PhysRevD.90.084035}{{\em Phys. Rev. D}
  {\bfseries 90} no.~8, (2014) 084035},
  \href{http://arxiv.org/abs/1406.6987}{{\ttfamily arXiv:1406.6987 [hep-th]}}.

\bibitem{Adler:1964um}
S.~L. Adler, ``{Consistency conditions on the strong interactions implied by a
  partially conserved axial vector current},''
  \href{http://dx.doi.org/10.1103/PhysRev.137.B1022}{{\em Phys. Rev.}
  {\bfseries 137} (1965) B1022--B1033}.

\bibitem{Meetz:1969as}
K.~Meetz, ``{Realization of chiral symmetry in a curved isospin space},''
  \href{http://dx.doi.org/10.1063/1.1664881}{{\em J. Math. Phys.} {\bfseries
  10} (1969) 589--593}.

\bibitem{Honerkamp:1971sh}
J.~Honerkamp, ``{Chiral multiloops},''
  \href{http://dx.doi.org/10.1016/0550-3213(72)90299-4}{{\em Nucl. Phys. B}
  {\bfseries 36} (1972) 130--140}.

\bibitem{Honerkamp:1971xtx}
J.~Honerkamp and K.~Meetz, ``{Chiral-invariant perturbation theory},''
  \href{http://dx.doi.org/10.1103/PhysRevD.3.1996}{{\em Phys. Rev. D}
  {\bfseries 3} (1971) 1996--1998}.

\bibitem{Ecker:1971xko}
G.~Ecker and J.~Honerkamp, ``{Application of invariant renormalization to the
  nonlinear chiral invariant pion lagrangian in the one-loop approximation},''
  \href{http://dx.doi.org/10.1016/0550-3213(71)90468-8}{{\em Nucl. Phys. B}
  {\bfseries 35} (1971) 481--492}.

\bibitem{Alvarez-Gaume:1981exa}
L.~Alvarez-Gaume, D.~Z. Freedman, and S.~Mukhi, ``{The Background Field Method
  and the Ultraviolet Structure of the Supersymmetric Nonlinear Sigma Model},''
  \href{http://dx.doi.org/10.1016/0003-4916(81)90006-3}{{\em Annals Phys.}
  {\bfseries 134} (1981) 85}.

\bibitem{Alvarez-Gaume:1981exv}
L.~Alvarez-Gaume and D.~Z. Freedman, ``{Geometrical Structure and Ultraviolet
  Finiteness in the Supersymmetric Sigma Model},''
  \href{http://dx.doi.org/10.1007/BF01208280}{{\em Commun. Math. Phys.}
  {\bfseries 80} (1981) 443}.

\bibitem{Boulware:1981ns}
D.~G. Boulware and L.~S. Brown, ``{Symmetric Space Scalar Field Theory},''
  \href{http://dx.doi.org/10.1016/0003-4916(82)90192-0}{{\em Annals Phys.}
  {\bfseries 138} (1982) 392}.

\bibitem{Howe:1986vm}
P.~S. Howe, G.~Papadopoulos, and K.~S. Stelle, ``{The Background Field Method
  and the Nonlinear $\sigma$ Model},''
  \href{http://dx.doi.org/10.1016/0550-3213(88)90379-3}{{\em Nucl. Phys. B}
  {\bfseries 296} (1988) 26--48}.

\bibitem{Dixon:1989fj}
L.~J. Dixon, V.~Kaplunovsky, and J.~Louis, ``{On Effective Field Theories
  Describing (2,2) Vacua of the Heterotic String},''
  \href{http://dx.doi.org/10.1016/0550-3213(90)90057-K}{{\em Nucl. Phys. B}
  {\bfseries 329} (1990) 27--82}.

\bibitem{Alonso:2015fsp}
R.~Alonso, E.~E. Jenkins, and A.~V. Manohar, ``{A Geometric Formulation of
  Higgs Effective Field Theory: Measuring the Curvature of Scalar Field
  Space},'' \href{http://dx.doi.org/10.1016/j.physletb.2016.01.041}{{\em Phys.
  Lett. B} {\bfseries 754} (2016) 335--342},
  \href{http://arxiv.org/abs/1511.00724}{{\ttfamily arXiv:1511.00724
  [hep-ph]}}.

\bibitem{Alonso:2016oah}
R.~Alonso, E.~E. Jenkins, and A.~V. Manohar, ``{Geometry of the Scalar
  Sector},'' \href{http://dx.doi.org/10.1007/JHEP08(2016)101}{{\em JHEP}
  {\bfseries 08} (2016) 101}, \href{http://arxiv.org/abs/1605.03602}{{\ttfamily
  arXiv:1605.03602 [hep-ph]}}.

\bibitem{Nagai:2019tgi}
R.~Nagai, M.~Tanabashi, K.~Tsumura, and Y.~Uchida, ``{Symmetry and geometry in
  a generalized Higgs effective field theory: Finiteness of oblique corrections
  versus perturbative unitarity},''
  \href{http://dx.doi.org/10.1103/PhysRevD.100.075020}{{\em Phys. Rev. D}
  {\bfseries 100} no.~7, (2019) 075020},
  \href{http://arxiv.org/abs/1904.07618}{{\ttfamily arXiv:1904.07618
  [hep-ph]}}.

\bibitem{Cohen:2021ucp}
T.~Cohen, N.~Craig, X.~Lu, and D.~Sutherland, ``{Unitarity Violation and the
  Geometry of Higgs EFTs},'' \href{http://arxiv.org/abs/2108.03240}{{\ttfamily
  arXiv:2108.03240 [hep-ph]}}.

\bibitem{Lehmann:1954rq}
H.~Lehmann, K.~Symanzik, and W.~Zimmermann, ``{On the formulation of quantized
  field theories},'' \href{http://dx.doi.org/10.1007/BF02731765}{{\em Nuovo
  Cim.} {\bfseries 1} (1955) 205--225}.

\bibitem{Kampf:2019mcd}
K.~Kampf, J.~Novotny, M.~Shifman, and J.~Trnka, ``{New Soft Theorems for
  Goldstone Boson Amplitudes},''
  \href{http://dx.doi.org/10.1103/PhysRevLett.124.111601}{{\em Phys. Rev.
  Lett.} {\bfseries 124} no.~11, (2020) 111601},
  \href{http://arxiv.org/abs/1910.04766}{{\ttfamily arXiv:1910.04766
  [hep-th]}}.

\bibitem{Camporesi:1990wm}
R.~Camporesi, ``{Harmonic analysis and propagators on homogeneous spaces},''
  \href{http://dx.doi.org/10.1016/0370-1573(90)90120-Q}{{\em Phys. Rept.}
  {\bfseries 196} (1990) 1--134}.

\bibitem{KobayashiNomizu}
S.~Kobayashi and K.~Nomizu, {\em Foundations of Differential Geometry, Volume
  2}.
\newblock A Wiley Publication in Applied Statistics. Wiley, 1996.

\bibitem{Braaten:1985is}
E.~Braaten, T.~L. Curtright, and C.~K. Zachos, ``{Torsion and Geometrostasis in
  Nonlinear Sigma Models},''
  \href{http://dx.doi.org/10.1016/0550-3213(86)90196-3}{{\em Nucl. Phys. B}
  {\bfseries 260} (1985) 630}. [Erratum: Nucl.Phys.B 266, 748--748 (1986)].

\bibitem{Nomizu}
K.~Nomizu, ``Invariant affine connections on homogeneous spaces,'' {\em
  American Journal of Mathematics} {\bfseries 76} no.~1, (1954) 33--65.
  \url{http://www.jstor.org/stable/2372398}.

\bibitem{Helgason}
S.~Helgason, {\em Differential Geometry, Lie Groups, and Symmetric Spaces}.
\newblock Crm Proceedings \& Lecture Notes. American Mathematical Society,
  2001.

\bibitem{Low:2014nga}
I.~Low, ``{Adler\textquoteright{}s zero and effective Lagrangians for
  nonlinearly realized symmetry},''
  \href{http://dx.doi.org/10.1103/PhysRevD.91.105017}{{\em Phys. Rev. D}
  {\bfseries 91} no.~10, (2015) 105017},
  \href{http://arxiv.org/abs/1412.2145}{{\ttfamily arXiv:1412.2145 [hep-th]}}.

\bibitem{Cheung:2020tqz}
C.~Cheung and Z.~Moss, ``{Symmetry and Unification from Soft Theorems and
  Unitarity},'' \href{http://dx.doi.org/10.1007/JHEP05(2021)161}{{\em JHEP}
  {\bfseries 05} (2021) 161}, \href{http://arxiv.org/abs/2012.13076}{{\ttfamily
  arXiv:2012.13076 [hep-th]}}.

\bibitem{Rodina:2021isd}
L.~Rodina and Z.~Yin, ``{Exploring the Landscape for Soft Theorems of Nonlinear
  Sigma Models},'' \href{http://arxiv.org/abs/2102.08396}{{\ttfamily
  arXiv:2102.08396 [hep-th]}}.

\bibitem{Kleiss:1988ne}
R.~Kleiss and H.~Kuijf, ``{Multi - Gluon Cross-sections and Five Jet Production
  at Hadron Colliders},''
  \href{http://dx.doi.org/10.1016/0550-3213(89)90574-9}{{\em Nucl. Phys. B}
  {\bfseries 312} (1989) 616--644}.

\bibitem{DelDuca:1999rs}
V.~Del~Duca, L.~J. Dixon, and F.~Maltoni, ``{New color decompositions for gauge
  amplitudes at tree and loop level},''
  \href{http://dx.doi.org/10.1016/S0550-3213(99)00809-3}{{\em Nucl. Phys. B}
  {\bfseries 571} (2000) 51--70},
  \href{http://arxiv.org/abs/hep-ph/9910563}{{\ttfamily arXiv:hep-ph/9910563}}.

\bibitem{Kawai:1985xq}
H.~Kawai, D.~C. Lewellen, and S.~H.~H. Tye, ``{A Relation Between Tree
  Amplitudes of Closed and Open Strings},''
  \href{http://dx.doi.org/10.1016/0550-3213(86)90362-7}{{\em Nucl. Phys. B}
  {\bfseries 269} (1986) 1--23}.

\bibitem{Bern:2008qj}
Z.~Bern, J.~J.~M. Carrasco, and H.~Johansson, ``{New Relations for Gauge-Theory
  Amplitudes},'' \href{http://dx.doi.org/10.1103/PhysRevD.78.085011}{{\em Phys.
  Rev. D} {\bfseries 78} (2008) 085011},
  \href{http://arxiv.org/abs/0805.3993}{{\ttfamily arXiv:0805.3993 [hep-ph]}}.

\bibitem{Bern:2010ue}
Z.~Bern, J.~J.~M. Carrasco, and H.~Johansson, ``{Perturbative Quantum Gravity
  as a Double Copy of Gauge Theory},''
  \href{http://dx.doi.org/10.1103/PhysRevLett.105.061602}{{\em Phys. Rev.
  Lett.} {\bfseries 105} (2010) 061602},
  \href{http://arxiv.org/abs/1004.0476}{{\ttfamily arXiv:1004.0476 [hep-th]}}.

\bibitem{Bern:2019prr}
Z.~Bern, J.~J. Carrasco, M.~Chiodaroli, H.~Johansson, and R.~Roiban, ``{The
  Duality Between Color and Kinematics and its Applications},''
  \href{http://arxiv.org/abs/1909.01358}{{\ttfamily arXiv:1909.01358
  [hep-th]}}.

\bibitem{Eschenburg}
J.-H. Eschenburg, ``Lecture notes on symmetric spaces.''
\newblock \url{http://myweb.rz.uni-augsburg.de/~eschenbu/symspace.pdf}.
  [Online; 20/10/2021].

\bibitem{Callan:1970yg}
C.~G. Callan, Jr., ``{Broken scale invariance in scalar field theory},''
  \href{http://dx.doi.org/10.1103/PhysRevD.2.1541}{{\em Phys. Rev. D}
  {\bfseries 2} (1970) 1541--1547}.

\bibitem{Boels:2015pta}
R.~H. Boels and W.~Wormsbecher, ``{Spontaneously broken conformal invariance in
  observables},'' \href{http://arxiv.org/abs/1507.08162}{{\ttfamily
  arXiv:1507.08162 [hep-th]}}.

\bibitem{Huang:2015sla}
Y.-t. Huang and C.~Wen, ``{Soft theorems from anomalous symmetries},''
  \href{http://dx.doi.org/10.1007/JHEP12(2015)143}{{\em JHEP} {\bfseries 12}
  (2015) 143}, \href{http://arxiv.org/abs/1509.07840}{{\ttfamily
  arXiv:1509.07840 [hep-th]}}.

\bibitem{DiVecchia:2015jaq}
P.~Di~Vecchia, R.~Marotta, M.~Mojaza, and J.~Nohle, ``{New soft theorems for
  the gravity dilaton and the Nambu-Goldstone dilaton at subsubleading
  order},'' \href{http://dx.doi.org/10.1103/PhysRevD.93.085015}{{\em Phys. Rev.
  D} {\bfseries 93} no.~8, (2016) 085015},
  \href{http://arxiv.org/abs/1512.03316}{{\ttfamily arXiv:1512.03316
  [hep-th]}}.

\bibitem{DiVecchia:2017uqn}
P.~Di~Vecchia, R.~Marotta, and M.~Mojaza, ``{Double-soft behavior of the
  dilaton of spontaneously broken conformal invariance},''
  \href{http://dx.doi.org/10.1007/JHEP09(2017)001}{{\em JHEP} {\bfseries 09}
  (2017) 001}, \href{http://arxiv.org/abs/1705.06175}{{\ttfamily
  arXiv:1705.06175 [hep-th]}}.

\bibitem{Weinberg:1966kf}
S.~Weinberg, ``{Pion scattering lengths},''
  \href{http://dx.doi.org/10.1103/PhysRevLett.17.616}{{\em Phys. Rev. Lett.}
  {\bfseries 17} (1966) 616--621}.

\bibitem{Arkani-Hamed:2008owk}
N.~Arkani-Hamed, F.~Cachazo, and J.~Kaplan, ``{What is the Simplest Quantum
  Field Theory?},'' \href{http://dx.doi.org/10.1007/JHEP09(2010)016}{{\em JHEP}
  {\bfseries 09} (2010) 016}, \href{http://arxiv.org/abs/0808.1446}{{\ttfamily
  arXiv:0808.1446 [hep-th]}}.

\bibitem{Cachazo:2015ksa}
F.~Cachazo, S.~He, and E.~Y. Yuan, ``{New Double Soft Emission Theorems},''
  \href{http://dx.doi.org/10.1103/PhysRevD.92.065030}{{\em Phys. Rev. D}
  {\bfseries 92} no.~6, (2015) 065030},
  \href{http://arxiv.org/abs/1503.04816}{{\ttfamily arXiv:1503.04816
  [hep-th]}}.

\bibitem{Du:2015esa}
Y.-J. Du and H.~Luo, ``{On single and double soft behaviors in NLSM},''
  \href{http://dx.doi.org/10.1007/JHEP08(2015)058}{{\em JHEP} {\bfseries 08}
  (2015) 058}, \href{http://arxiv.org/abs/1505.04411}{{\ttfamily
  arXiv:1505.04411 [hep-th]}}.

\bibitem{Low:2015ogb}
I.~Low, ``{Double Soft Theorems and Shift Symmetry in Nonlinear Sigma
  Models},'' \href{http://dx.doi.org/10.1103/PhysRevD.93.045032}{{\em Phys.
  Rev. D} {\bfseries 93} no.~4, (2016) 045032},
  \href{http://arxiv.org/abs/1512.01232}{{\ttfamily arXiv:1512.01232
  [hep-th]}}.

\bibitem{Susskind:1970gf}
L.~Susskind and G.~Frye, ``{Algebraic aspects of pionic duality diagrams},''
  \href{http://dx.doi.org/10.1103/PhysRevD.1.1682}{{\em Phys. Rev. D}
  {\bfseries 1} (1970) 1682--1686}.

\bibitem{Osborn:1969ku}
H.~Osborn, ``{Implications of adler zeros for multipion processes},''
  \href{http://dx.doi.org/10.1007/BF02755724}{{\em Lett. Nuovo Cim.} {\bfseries
  2S1} (1969) 717--723}.

\bibitem{Cheung:2014dqa}
C.~Cheung, K.~Kampf, J.~Novotny, and J.~Trnka, ``{Effective Field Theories from
  Soft Limits of Scattering Amplitudes},''
  \href{http://dx.doi.org/10.1103/PhysRevLett.114.221602}{{\em Phys. Rev.
  Lett.} {\bfseries 114} no.~22, (2015) 221602},
  \href{http://arxiv.org/abs/1412.4095}{{\ttfamily arXiv:1412.4095 [hep-th]}}.

\bibitem{Cheung:2016drk}
C.~Cheung, K.~Kampf, J.~Novotny, C.-H. Shen, and J.~Trnka, ``{A Periodic Table
  of Effective Field Theories},''
  \href{http://dx.doi.org/10.1007/JHEP02(2017)020}{{\em JHEP} {\bfseries 02}
  (2017) 020}, \href{http://arxiv.org/abs/1611.03137}{{\ttfamily
  arXiv:1611.03137 [hep-th]}}.

\bibitem{Hinterbichler:2015pqa}
K.~Hinterbichler and A.~Joyce, ``{Hidden symmetry of the Galileon},''
  \href{http://dx.doi.org/10.1103/PhysRevD.92.023503}{{\em Phys. Rev. D}
  {\bfseries 92} no.~2, (2015) 023503},
  \href{http://arxiv.org/abs/1501.07600}{{\ttfamily arXiv:1501.07600
  [hep-th]}}.

\bibitem{Cheung:2015ota}
C.~Cheung, K.~Kampf, J.~Novotny, C.-H. Shen, and J.~Trnka, ``{On-Shell
  Recursion Relations for Effective Field Theories},''
  \href{http://dx.doi.org/10.1103/PhysRevLett.116.041601}{{\em Phys. Rev.
  Lett.} {\bfseries 116} no.~4, (2016) 041601},
  \href{http://arxiv.org/abs/1509.03309}{{\ttfamily arXiv:1509.03309
  [hep-th]}}.

\bibitem{Kampf:2013vha}
K.~Kampf, J.~Novotny, and J.~Trnka, ``{Tree-level Amplitudes in the Nonlinear
  Sigma Model},'' \href{http://dx.doi.org/10.1007/JHEP05(2013)032}{{\em JHEP}
  {\bfseries 05} (2013) 032}, \href{http://arxiv.org/abs/1304.3048}{{\ttfamily
  arXiv:1304.3048 [hep-th]}}.

\bibitem{Luo:2015tat}
H.~Luo and C.~Wen, ``{Recursion relations from soft theorems},''
  \href{http://dx.doi.org/10.1007/JHEP03(2016)088}{{\em JHEP} {\bfseries 03}
  (2016) 088}, \href{http://arxiv.org/abs/1512.06801}{{\ttfamily
  arXiv:1512.06801 [hep-th]}}.

\bibitem{Elvang:2017mdq}
H.~Elvang, M.~Hadjiantonis, C.~R.~T. Jones, and S.~Paranjape, ``{On the
  Supersymmetrization of Galileon Theories in Four Dimensions},''
  \href{http://dx.doi.org/10.1016/j.physletb.2018.04.032}{{\em Phys. Lett. B}
  {\bfseries 781} (2018) 656--663},
  \href{http://arxiv.org/abs/1712.09937}{{\ttfamily arXiv:1712.09937
  [hep-th]}}.

\bibitem{Elvang:2018dco}
H.~Elvang, M.~Hadjiantonis, C.~R.~T. Jones, and S.~Paranjape, ``{Soft Bootstrap
  and Supersymmetry},'' \href{http://dx.doi.org/10.1007/JHEP01(2019)195}{{\em
  JHEP} {\bfseries 01} (2019) 195},
  \href{http://arxiv.org/abs/1806.06079}{{\ttfamily arXiv:1806.06079
  [hep-th]}}.

\bibitem{Cheung:2018oki}
C.~Cheung, K.~Kampf, J.~Novotny, C.-H. Shen, J.~Trnka, and C.~Wen, ``{Vector
  Effective Field Theories from Soft Limits},''
  \href{http://dx.doi.org/10.1103/PhysRevLett.120.261602}{{\em Phys. Rev.
  Lett.} {\bfseries 120} no.~26, (2018) 261602},
  \href{http://arxiv.org/abs/1801.01496}{{\ttfamily arXiv:1801.01496
  [hep-th]}}.

\bibitem{Low:2019ynd}
I.~Low and Z.~Yin, ``{Soft Bootstrap and Effective Field Theories},''
  \href{http://dx.doi.org/10.1007/JHEP11(2019)078}{{\em JHEP} {\bfseries 11}
  (2019) 078}, \href{http://arxiv.org/abs/1904.12859}{{\ttfamily
  arXiv:1904.12859 [hep-th]}}.

\bibitem{Kampf:2021bet}
K.~Kampf, J.~Novotny, F.~Preucil, and J.~Trnka, ``{Multi-spin soft bootstrap
  and scalar-vector Galileon},''
  \href{http://arxiv.org/abs/2104.10693}{{\ttfamily arXiv:2104.10693
  [hep-th]}}.

\bibitem{Kampf:2021tbk}
K.~Kampf, J.~Novotny, and P.~Vasko, ``{Extended DBI and its generalizations
  from graded soft theorems},''
  \href{http://dx.doi.org/10.1007/JHEP10(2021)101}{{\em JHEP} {\bfseries 10}
  (2021) 101}, \href{http://arxiv.org/abs/2107.04587}{{\ttfamily
  arXiv:2107.04587 [hep-th]}}.

\bibitem{Britto:2004ap}
R.~Britto, F.~Cachazo, and B.~Feng, ``{New recursion relations for tree
  amplitudes of gluons},''
  \href{http://dx.doi.org/10.1016/j.nuclphysb.2005.02.030}{{\em Nucl. Phys. B}
  {\bfseries 715} (2005) 499--522},
  \href{http://arxiv.org/abs/hep-th/0412308}{{\ttfamily arXiv:hep-th/0412308}}.

\bibitem{Britto:2005fq}
R.~Britto, F.~Cachazo, B.~Feng, and E.~Witten, ``{Direct proof of tree-level
  recursion relation in Yang-Mills theory},''
  \href{http://dx.doi.org/10.1103/PhysRevLett.94.181602}{{\em Phys. Rev. Lett.}
  {\bfseries 94} (2005) 181602},
  \href{http://arxiv.org/abs/hep-th/0501052}{{\ttfamily arXiv:hep-th/0501052}}.

\end{thebibliography}\endgroup

\end{document}